\documentclass[12pt]{article}
\usepackage[latin9]{inputenc}
\usepackage{color}
\usepackage{array}
\usepackage{float}
\usepackage{units}
\usepackage{multirow}
\usepackage{amsmath}
\usepackage{amssymb}
\usepackage{cancel}
\usepackage{graphicx}
\usepackage{rotfloat}

\makeatletter

\providecommand{\tabularnewline}{\\}

\usepackage{ulem}
\usepackage{cite}
\usepackage{rotating}
\usepackage{amsfonts}
\usepackage{multirow}

\newcommand{\beq}{\begin{equation}}
\newcommand{\eeq}{\end{equation}}
\newcommand{\ri}{\mathrm{i}}

\makeatother

\begin{document}
\begin{titlepage}

\vskip 2cm 
\begin{center}
\textbf{\Large{}{}{}{}Non-commutativity and non-inertial effects
on a scalar field in a cosmic string space-time}{\Large{}{}}\footnote{\texttt{Email: rodrigo.cuzinatto@unifal-mg.edu.br, mdemonti@ualberta.ca,
pompeia@ita.br}}{\Large{}{}}\\
 {\Large{}{} {} }\textbf{\Large{}{}{}{}Part 1: Klein-Gordon oscillator}{\Large\par}
\par\end{center}

\begin{center} 
\par\end{center}

\begin{center}
\textbf{\Large{}}\textbf{ \vskip 10pt Rodrigo Rocha Cuzinatto$^{a}$, Marc de Montigny$^{b}$,
Pedro Jos\'e Pompeia$^{c}$ } \vskip 5pt \textsl{$^{a}$Instituto de
Ci\^encia e Tecnologia, Universidade Federal de Alfenas}\\
\textsl{ Rodovia Jos\'e Aur\'elio Vilela, 11999, Cidade Universit\'aria}\\
\textsl{ CEP 37715-400 Po\c cos de Caldas, Minas Gerais, Brazil} \vskip
2pt \textsl{$^{b}$Facult\'e Saint-Jean, University of Alberta }\\
\textsl{ 8406 91 Street NW}\\
\textsl{ Edmonton, Alberta, Canada T6C 4G9, Canada} \vskip 2pt \textsl{$^{c}$Departamento
de F\'{\i}sica, Instituto Tecnol\'ogico de Aeron\'autica}\\
\textsl{ Pra\c ca Mal. Eduardo Gomes 50}\\
\textsl{ CEP 12228-900 S\~ao Jos\'e dos Campos, S\~ao Paulo, Brazil} 
\par\end{center}

\begin{abstract}
We analyse the Klein-Gordon oscillator in a cosmic string space-time
and study the effects stemming from the rotating frame and non-commutativity
in momentum space. We show that the latter mimics a constant magnetic
field, imparting physical interpretation to the setup. The field equation
for the scalar field is solved via separations of variables, and we
obtain quantization of energy and angular momentum. The space-time
metric is non-degenerate as long as the particle is confined within
a hard-wall, whose position depends on the rotation frame velocity
and the string mass parameter. We investigate the energy quantization
both for a finite hard-wall (numerical evaluation) and in the limit
of an infinite hard-wall (analytical treatment). We stress the effect
of non-commutativity upon the energy quantization in each case. 
\end{abstract}
\bigskip{}


\end{titlepage}

\newpage{}

\section{Introduction{\label{introduction}}}

The hypothetical cosmological objects known as cosmic strings are
among the most-investigated examples of topological defect space-times
which have stimulated interest in theoretical physics for more than
forty years \cite{Kibble1976,6,5,VilenkinShellard}. In particular,
the gravitational properties of cosmic strings are remarkably different
from typical linear distributions of matter. For instance, the geometry
of the space-time around a straight cosmic string is locally identical
to a flat space-time, but globally conical rather than Euclidean,
and with an azimuthal deficit angle related to the string tension
\cite{Vilenkin1981,VilenkinShellard}. Naturally an interesting topic
is the interaction between such gravitational fields and quantum fields.
The Klein-Gordon (KG) scalar field is among the most basic cases to
examine in that context. The KG equation in various types of curved
space-times was discussed, e.g. in Refs. \cite{AhmedGERG2019,VitBak2018,VitBakIJMPD2018,Barros2018,VitBak2016,Cunha2016,BakFurAP2015,MotaBak2014,BoumaliCJP2014,Medeiros2012,Caval2006,Yang2021,Zhong2021,MdM2021}.
In particular, the generalized Klein-Gordon oscillator has been recently
studied in a G\"odel-type space-time \cite{Yang2021}, Som-Raychaudhuri
space-time \cite{Zhong2021} and in a global monopole space-time \cite{MdM2021}.

In this paper, we analyze the effects of non-commutativity on the
relativistic dynamics of a KG oscillator in a rotating cosmic string
space-time. In a recent paper, we studied the analogous problem for
fermions by examining the non-inertial effects of a rotating frame
on a Dirac oscillator in a cosmic string space-time with non-commutative
geometry in the phase space \cite{GERG2019}. The KG equation describes
spin-zero fields, whether they are composite particles, like pions,
or the Higgs boson, which is the first known elementary particle with
spin-zero. The so-called `KG oscillator' is defined in a way analogous
to the `Dirac oscillator', which was introduced in 1989 \cite{Moshinsky,Ito,Cook}
and represents an exactly solvable model of many-particle models in
relativistic quantum mechanics.

Hereafter we investigate the interaction between the KG oscillator
and cosmic strings, which may be seen as spatial lines of trapped
energy density, in analogy with vortex lines in superfluids and superconductors
or line defects in crystals \cite{Brandenberger2014,5,VilenkinShellard,Hindmarsh}.
Our main interest here is to evaluate the effects of non-commutativity
in the phase space on the KG equation's wave functions and energy
eigenvalues. Non-commutative geometry has been applied recently to
the quantum Hall effect \cite{Hall,Hall2,Hall3,Hall4}, geometric
phases \cite{Passos,Ribeiro}, the Dirac oscillator \cite{Mandal2,Panella,Cai,Zarrinkamar,Hou}
and the relativistic Duffin-Kemmer-Petiau (DKP) oscillator \cite{Yang,Guo,Falek}.
The DKP approach aims at treating integer-spin fields with first-order
field equations in a way that resembles the Dirac equation for half-integer
spin \cite{Duffin,Kemmer,Petiau,Greiner}. The spin-zero DKP equation
can be shown to the equivalent to KG equation in the case of free
fields \cite{Lunardi}. However, this equivalence does not necessarily
hold when interactions come into play \cite{Guertin,VixeMaria}. In
fact, gauge theory taught us long ago \cite{Utiyama} that interactions
can be introduced by means of a coupling prescription. The `DKP oscillator'
is built from a prescription that is different from the coupling prescription
characterizing the KG oscillator. This fact will naturaly lead to
different physical consequences in each case. This paper explores
the physical significance of KG oscillator in the presence of non-commutativity,
rotating frame and a cosmic string ambient space-time; the companion
paper \cite{NCSF_Part2} scrutinizes DKP oscillator for the spin-zero
field in an analogous context.

Here, we consider a cosmic string space-time equipped with a rotating
frame in cylindrical coordinates. We work with the cosmic string space-time
with metric signature diag$(-1,+1,+1,+1)$, and described in cylindrical
coordinates $(\rho,\varphi,z)$ where $\rho\in[0,\infty),\,\varphi\in[0,2\pi],\,z\in(-\infty,+\infty)$
by the line element 
\begin{equation}
ds^{2}=-\left(1-\omega^{2}\eta^{2}\rho^{2}\right)dt^{2}+2\omega\eta^{2}\rho^{2}d\varphi dt+d\rho^{2}+\eta^{2}\rho^{2}d\varphi^{2}+dz^{2}.\label{CosmicString}
\end{equation}
In Eq. (\ref{CosmicString}), $\rho=\sqrt{x^{2}+y^{2}}$, $\omega$
is the angular frequency of the rotating frame and $\eta=1-4\Lambda\in(0,1]$
(with $\Lambda$ the string's linear mass density) is related to the
deficit angle $\theta=2\pi\left(1-\eta\right)$. We consider units
such that the speed of light $c=1$ and the Planck's constant $\hbar=1$.
The line element in Eq. (\ref{CosmicString}) corresponds geometrically
to a Minkowski space-time with a conical singularity \cite{deMello2004}.
From the first term of Eq. (\ref{CosmicString}) we deduce two natural
intervals, delimited by 
\begin{equation}
\rho_{0}\equiv\frac{1}{\omega\eta},\qquad0<\rho<\rho_{0}\ {\mathrm{and}}\ \rho>\rho_{0}.\label{hardwall}
\end{equation}
For the latter, the particle lies outside of the light-cone since
its speed is greater than the speed of light, which implies that the
wave function of the particle must vanish as $\rho$ approaches $\rho_{0}$.
Thus $\omega\eta$ determines two classes of solutions: firstly, a
finite wall $\rho_{0}$ with $\omega\eta$ arbitrary but finite, and
secondly, a wall $\rho_{0}$ at infinity when $\omega\eta\ll1$ \cite{Bakke}.
We will say more about this in Section \ref{Energy}.

In this paper, we examine the effects of non-commutativity in a cosmic
string space-time with a rotating frame in cylindrical coordinates
by solving the KG equation in this curved space-time. The geometry
of the spacetime can play the role of a hard-wall confining potential
via non-inertial effects (see Ref. \cite{Bakke} and references therein).
Hereafter, for simplicity, we consider non-commutativity in the momentum
components only, and we restrict the non-commutativity parameters
to be parallel to the cosmic string. Our purpose is to find the solutions
of the field equations and the corresponding energies. In Section
\ref{SecCosmicString}, we solve the KG equation and find solutions
restricted to the plane $z=0$ as well as general solutions in terms
of $z$. In Section \ref{subsec:MagField} we show that the non-commutativity
parameter can be interpreted as a magnetic field. In Section \ref{Energy},
we obtain the energy eigenvalues which correspond to the wave functions
and discuss the dissipative effects arising from the coupling prescription.
Finally, in Section \ref{Conclusion}, we discuss our results and
propose potential applications.




\section{Klein-Gordon oscillator in a non-inertial frame and cosmic string
space-time with non-commutative geometry{\label{SecCosmicString}}}

The equation for the KG oscillator is obtained from the KG equation
by applying the non-minimal coupling \cite{Moshinsky,Bakke,Nedjadi}:
\begin{equation}
\mathbf{p}\phi\ {\rightarrow}\ \left(\mathbf{p}-im\omega_{0}\mathbf{r}\right)\phi,\qquad\mathbf{p}\phi^{*}\ {\rightarrow}\ \left(\mathbf{p}+im\omega_{0}\mathbf{r}\right)\phi^{*},\label{nonminimal}
\end{equation}
where we carefully apply the appropriate sign for the complex conjugate
field. An alternative possibility for the coupling prescription would
be to keep the same sign for both $\mathbf{p}\phi$ and $\mathbf{p}\phi^{*}$;
the motivation for doing that and its consequences is explored in
Part 2 of this work \cite{NCSF_Part2}. The KG field has a mass $m$
and the oscillator's frequency is denoted by $\omega_{0}$. Hereafter,
we use cylindrical coordinates and select 
\begin{equation}
\mathbf{r}=\rho\hat{e}_{\rho}+z\hat{e}_{z}=\left(\rho,0,z\right).
\end{equation}
Since $p_{j}=-{\ri}\partial_{j}$ ($j=1,2,3$, standing respectively
for $\rho$, $\varphi$ and $z$), we can express Eq. (\ref{nonminimal})
as $-\ri\left(\nabla\pm m\omega_{0}\mathbf{r}\right)$, where $+$
(resp. $-$) applies to $\phi$ (resp. $\phi^{*}$).

If we begin with the KG Lagrangian 
\begin{equation}
L=\sqrt{-g}\left(-g^{ab}\partial_{a}\phi^{*}\partial_{b}\phi-m^{2}\phi^{*}\phi\right),
\end{equation}
and perform the non-minimal substitution in Eq. \eqref{nonminimal},
we obtain the Lagrangian of the KG oscillator: 
\begin{eqnarray}
L_{\mathrm{osc}} & = & -\sqrt{-g}\left[g^{00}\partial_{0}\phi^{*}\partial_{0}\phi+g^{02}\partial_{0}\phi^{*}\partial_{2}\phi+g^{02}\partial_{2}\phi^{*}\partial_{0}\phi\right.\nonumber \\
 &  & +g^{11}\left(\partial_{1}\phi^{*}-m\omega_{0}\rho\phi^{*}\right)\left(\partial_{1}\phi+m\omega_{0}\rho\phi\right)+g^{22}\partial_{2}\phi^{*}\partial_{2}\phi\label{Lag-osc}\\
 &  & \left.+g^{33}\left(\partial_{3}\phi^{*}-m\omega_{0}z\phi^{*}\right)\left(\partial_{3}\phi+m\omega_{0}z\phi\right)+m^{2}\phi^{*}\phi\right],\nonumber 
\end{eqnarray}
which, with 0, 1, 2, 3 standing for $t$, $\rho$, $\varphi$ and
$z$, respectively, leads to the equation of motion 
\begin{eqnarray}
-m^{2}\phi+m^{2}\omega_{0}^{2}\rho^{2}\phi+m^{2}\omega_{0}^{2}z^{2}\phi-\partial_{0}^{2}\phi+\left(2m\omega_{0}\rho+\frac{1}{\rho}\right)\partial_{1}\phi+\partial_{1}^{2}\phi\nonumber \\
+\left(\frac{1}{\eta^{2}\rho^{2}}-\omega^{2}\right)\partial_{2}^{2}\phi+2\omega\partial_{2}\partial_{0}\phi+2m\omega_{0}z\partial_{3}\phi+\partial_{3}^{2}\phi+3m\omega_{0}\phi=0.\label{RodEq4}
\end{eqnarray}

Next we introduce the non-commutative momentum space, which is a particular
case from the general non-commutative phase space described by the
operators $\hat{r}_{i}$ and $\hat{p}_{i}$, defined via the following
generalized Bopp shift \cite{GERG2019,Fairlie}: 
\begin{eqnarray}
\hat{r}_{i} & = & r_{i}-\frac{\Theta_{ij}}{2\hbar}p_{j}=r_{i}+\frac{\left(\mathbf{\Theta}\times{\bf p}\right)_{i}}{2\hbar},\label{transformation1}\\
\hat{p}_{i} & = & p_{i}+\frac{\Omega_{ij}}{2\hbar}r_{j}=p_{i}-\frac{\left(\mathbf{\Omega}\times{\bf r}\right)_{i}}{2\hbar},\label{transformation2}
\end{eqnarray}
which satisfy the following commutation relations: 
\begin{eqnarray}
\left[\hat{r}_{i},\hat{r}_{j}\right]={\rm i}\Theta_{ij},\quad\left[\hat{p}_{i},\hat{p}_{j}\right]={\rm i}\Omega_{ij},\quad\left[\hat{r}_{i},\hat{p}_{j}\right]={\rm i}\hbar\Delta_{ij},
\end{eqnarray}
where $\Theta_{ij}=\epsilon_{ijk}\Theta_{k}$, $\Omega_{ij}=\epsilon_{ijk}\Omega_{k}$,
with $\Theta_{i}$ and $\Omega_{i}$ ($i=1,2,3$) real parameters.
The matrix $\Delta_{ij}$ is given by 
\begin{equation}
\Delta_{ij}=\left(1+\frac{\mathbf{\Theta}\cdot\mathbf{\Omega}}{4\hbar^{2}}\right)\delta_{ij}-\frac{\Omega_{i}\Theta_{j}}{4\hbar^{2}}.
\end{equation}

We will restrict our study to 
\begin{equation}
\Theta_{j}=0,\quad j=1,2,3,
\end{equation}
so that we allow for a non-commutative momentum space only, whereas
the configuration space for the cosmic string remains commutative.
The reason for this particular choice will be addressed at the end
of Section \ref{subsec:MagField}. Therefore, the non-commutative
phase-space components are 
\begin{align}
 & \hat{r}_{i}=r_{i},\nonumber \\
 & \hat{p}_{i}=p_{i}+\frac{\Omega_{ij}}{2}r_{j}=p_{i}-\frac{\left(\mathbf{\Omega}\times{\bf r}\right)_{i}}{2},\label{pNC}
\end{align}
where 
\begin{eqnarray}
\left[\hat{r}_{i},\hat{r}_{j}\right]=0,\quad\left[\hat{p}_{i},\hat{p}_{j}\right]={\rm i}\Omega_{ij},\quad\left[\hat{r}_{i},\hat{p}_{j}\right]={\rm i}\hbar\delta_{ij}.
\end{eqnarray}
In cylindrical coordinates, the second term of Eq. (\ref{pNC}) is
\begin{eqnarray}
\frac{\Omega_{ij}}{2}r_{j}=\frac{1}{2}\left(\Omega_{13}z,\Omega_{21}\rho+\Omega_{23}z,\Omega_{31}\rho\right)=\frac{1}{2}\left(-\Omega_{2}z,-\Omega_{3}\rho+\Omega_{1}z,\Omega_{2}\rho\right).
\end{eqnarray}

Then, Eq. (\ref{Lag-osc}) is generalized to the non-commutative oscillator
Lagrangian: 
\begin{eqnarray}
L_{\mathrm{NC-osc}} & = & -\sqrt{-g}\left[g^{00}\partial_{0}\phi^{*}\partial_{0}\phi+g^{02}\partial_{0}\phi^{*}\left(\partial_{2}\phi+i\frac{1}{2}\Omega_{3}\rho\phi-\frac{1}{2}i\Omega_{1}z\phi\right)\right. \nonumber \\
 &  & +g^{02}\left(\partial_{2}\phi^{*}-\frac{1}{2}i\Omega_{3}\rho\phi^{*}+\frac{1}{2}i\Omega_{1}z\phi^{*}\right)\partial_{0}\phi \nonumber \\
 &  & +g^{11}\left(\partial_{1}\phi^{*}-m\omega_{0}\rho\phi^{*}-\frac{1}{2}i\Omega_{2}z\phi^{*}\right)\left(\partial_{1}\phi+m\omega_{0}\rho\phi+\frac{1}{2}i\Omega_{2}z\phi\right) \nonumber \\
 &  & +g^{22}\left(\partial_{2}\phi^{*}-\frac{1}{2}i\Omega_{3}\rho\phi^{*}+\frac{1}{2}i\Omega_{1}z\phi^{*}\right)\left(\partial_{2}\phi+i\frac{1}{2}\Omega_{3}\rho\phi-\frac{1}{2}i\Omega_{1}z\phi\right) \nonumber \\
 &  & \left.+g^{33}\left(\partial_{3}\phi^{*}-m\omega_{0}z\phi^{*}+\frac{1}{2}i\Omega_{2}\rho\phi^{*}\right)\left(\partial_{3}\phi+m\omega_{0}z\phi-\frac{1}{2}i\Omega_{2}\rho\phi\right)+m^{2}\phi^{*}\phi\right],
\end{eqnarray}
which modifies Eq. (\ref{RodEq4}) as follows: 
\begin{equation}
\begin{array}{ccc}
-m^{2}\phi+3m\omega_{0}\phi+\left(m\omega_{0}\rho+i\frac{1}{2}\Omega_{2}z\right)^{2}\phi+\left(m\omega_{0}z-i\frac{1}{2}\Omega_{2}\rho\right)^{2}\phi\\
+\left(\frac{1}{\eta^{2}\rho^{2}}-\omega^{2}\right)\left(-i\frac{1}{2}\Omega_{3}\rho+i\frac{1}{2}\Omega_{1}z\right)^{2}\phi+\left(i\frac{1}{2}\Omega_{2}z\right)\frac{1}{\rho}\phi\\
-2\omega\left(-i\frac{1}{2}\Omega_{3}\rho+i\frac{1}{2}\Omega_{1}z\right)\partial_{0}\phi+2\left(m\omega_{0}\rho+i\frac{1}{2}\Omega_{2}z\right)\partial_{1}\phi\\
-2\left(\frac{1}{\eta^{2}\rho^{2}}-\omega^{2}\right)\left(-i\frac{1}{2}\Omega_{3}\rho+i\frac{1}{2}\Omega_{1}z\right)\partial_{2}\phi+2\left(m\omega_{0}z-i\frac{1}{2}\Omega_{2}\rho\right)\partial_{3}\phi\\
-\partial_{0}^{2}\phi+2\omega\partial_{0}\partial_{2}\phi+\partial_{1}^{2}\phi+\frac{1}{\rho}\partial_{1}\phi+\left(\frac{1}{\eta^{2}\rho^{2}}-\omega^{2}\right)\partial_{2}^{2}\phi+\partial_{3}^{2}\phi=0.
\end{array}\label{EMNC}
\end{equation}
The solution for this equation will be dealt with in Section \ref{subsec:Solution}.
Before that, we would like to address the mapping existing between
non-commutativity and a magnetic field.


\subsection{Non-commutativity as a magnetic field \label{subsec:MagField}}

The non-commutativity in momentum space, such as considering hereafter,
can be interpreted as a magnetic field. In this subsection, we first
analyze the \textit{commutative} Klein-Gordon oscillator in a cosmic
string space-time equipped with a rotating frame in the presence of
a magnetic field, and then map the equations on our non-commutative
model.

As usual we consider the minimal coupling between the KG and electromagnetic
fields, i.e. 
\begin{equation}
\mathbf{p}\rightarrow\mathbf{p}\mp e\mathbf{A}
\end{equation}
We perform this prescription at the level of the Lagrangian (differently
of the approach in \cite{Boumali}). This leads to: 
\begin{eqnarray}
L_{B-\mathrm{osc}} & = & -\sqrt{-g}\left[g^{00}\partial_{0}\phi^{*}\partial_{0}\phi+g^{02}\partial_{0}\phi^{*}\left(\partial_{2}\phi-ieA_{2}\phi\right)+g^{02}\left(\partial_{2}\phi^{*}+ieA_{2}\phi^{*}\right)\partial_{0}\phi\right. \nonumber \\
 &  & +g^{11}\left(\partial_{1}\phi^{*}-m\omega_{0}\rho\phi^{*}+ieA_{1}\phi^{*}\right)\left(\partial_{1}\phi+m\omega_{0}\rho\phi-ieA_{1}\phi\right) \nonumber \\
 &  & +g^{22}\left(\partial_{2}\phi^{*}+ieA_{2}\phi^{*}\right)\left(\partial_{2}\phi-ieA_{2}\phi\right) \nonumber \\
 &  & \left.+g^{33}\left(\partial_{3}\phi^{*}-m\omega_{0}z\phi^{*}+ieA_{3}\phi^{*}\right)\left(\partial_{3}\phi+m\omega_{0}z\phi-ieA_{3}\phi\right)+m^{2}\phi^{*}\phi\right].
\end{eqnarray}
From this Lagrangian we readily obtain the field equation: 
\begin{equation}
\begin{array}{ccc}
-m^{2}\phi+3m\omega_{0}\phi+\left(m\omega_{0}\rho-ieA_{1}\right)^{2}\phi+\left(m\omega_{0}z-ieA_{3}\right)^{2}\phi\\
+\left(\frac{1}{\eta^{2}\rho^{2}}-\omega^{2}\right)\left(ieA_{2}\right)^{2}\phi-\left(ieA_{1}\right)\frac{1}{\rho}\phi\\
-2\omega\left(ieA_{2}\right)\partial_{0}\phi+2\left(m\omega_{0}\rho-ieA_{1}\right)\partial_{1}\phi\\
-2\left(\frac{1}{\eta^{2}\rho^{2}}-\omega^{2}\right)\left(ieA_{2}\right)\partial_{2}\phi+2\left(m\omega_{0}z-ieA_{3}\right)\partial_{3}\phi\\
-\partial_{0}^{2}\phi+2\omega\partial_{0}\partial_{2}\phi+\partial_{1}^{2}\phi+\frac{1}{\rho}\partial_{1}\phi+\left(\frac{1}{\eta^{2}\rho^{2}}-\omega^{2}\right)\partial_{2}^{2}\phi+\partial_{3}^{2}\phi\\
-ie\left[\omega\partial_{0}A_{2}+\partial_{1}A_{1}+\left(\frac{1}{\eta^{2}\rho^{2}}-\omega^{2}\right)\partial_{2}A_{2}+\partial_{3}A_{3}\right]\phi=0.
\end{array}\label{EMMF}
\end{equation}

The comparison between the Lagrangians $L_{B-\mathrm{osc}}$ and $L_{\mathrm{NC-osc}}$
and the comparison between the field equations in Eqs. \eqref{EMNC}
and \eqref{EMMF} exhibit the following mapping between the non-commutativity
parameters and the components of vector potential: 
\begin{equation}
eA_{1}=-\frac{1}{2}\Omega_{2}z,\qquad eA_{2}=-\frac{1}{2}\Omega_{3}\rho+\frac{1}{2}\Omega_{1}z,\qquad eA_{3}=\frac{1}{2}\Omega_{2}\rho,
\end{equation}
so that $\partial_{1}A_{1}=\partial_{2}A_{2}=\partial_{0}A_{2}=\partial_{3}A_{3}=0$.
In particular, if we consider $\Omega_{1}=\Omega_{2}=0$, we are left
with 
\begin{align}
-e\nabla\times\mathbf{A} & =\Omega_{3}\hat{z},
\end{align}
which shows that our non-commutative parameter $\Omega_{3}$ represents
an external constant magnetic field along the $z$-direction as in
the Landau problem \cite{Li}. (This is also typical of a magnetic
field within an infinite solenoid.) As a second possibility, had we
considered $\Omega_{1}=\Omega_{3}=0$, we would have obtained: 
\begin{align}
-e\nabla\times\mathbf{A} & =\Omega_{2}\hat{\varphi}
\end{align}
which is a constant magnetic field in the $\hat{\varphi}$ direction,
representing a field that circles around the $z$-axis. (This might
describe the magnetic field inside a toroid whose inner radius is
much smaller than the outer radius.)

We will now analyze if non-commutativity in the coordinate space could
also be related to a magnetic field. For this, we take $\Omega_{j}=0$
but keep $\Theta_{j}\neq0$, $j=1,2,3$, so that the relations in
Eq. (\ref{pNC}) are replaced by 
\begin{align}
 & \hat{r}_{i}=r_{i}-\frac{\Theta_{ij}}{2\hbar}p_{j},=r_{i}+\frac{\left(\mathbf{\Theta}\times{\bf p}\right)_{i}}{2\hbar}, \nonumber \\
 & \hat{p}_{i}=p_{i}.
\end{align}
From this and Eq. \eqref{nonminimal}, the oscillator prescription
takes on the form 
\begin{equation}
{\bf p}\rightarrow\hat{{\bf p}}\mp{\rm i}m\omega_{0}\hat{{\bf r}}\rightarrow-{\rm i}\left(\nabla\pm m\omega_{0}{\bf r}\pm m\omega_{0}\frac{\left(\mathbf{\Theta}\times{\bf p}\right)}{2\hbar}\right),
\end{equation}
and the Lagrangian for the scalar field KG oscillator in the presence
of coordinate non-commutativity reads: 
\begin{eqnarray}
L_{r{\rm -NC-osc}} & = & -\sqrt{-g}\left[g^{00}\partial_{0}\phi^{*}\partial_{0}\phi+g^{02}\partial_{0}\phi^{*}\left(\partial_{2}\phi+\frac{1}{2}{\rm i}m\rho\omega_{0}\left[\Theta_{3}\partial_{\rho}\phi-\Theta_{1}\partial_{z}\phi\right]\right)\right. \nonumber \\
 &  & +g^{02}\left(\partial_{2}\phi^{*}-\frac{1}{2}{\rm i}m\rho\omega_{0}\left[\Theta_{3}\partial_{\rho}\phi^{*}-\Theta_{1}\partial_{z}\phi^{*}\right]\right)\partial_{0}\phi \nonumber \\
 &  & +g^{11}\left(\partial_{1}\phi^{*}-m\omega_{0}\rho\phi^{*}-\frac{1}{2}{\rm i}m\omega_{0}\left[-\Theta_{3}\frac{1}{\rho}\partial_{\varphi}\phi^{*}+\Theta_{2}\partial_{z}\phi^{*}\right]\right) \nonumber \\
 &  & \times\left(\partial_{1}\phi+m\omega_{0}\rho\phi+\frac{1}{2}{\rm i}m\omega_{0}\left[-\Theta_{3}\frac{1}{\rho}\partial_{\varphi}\phi+\Theta_{2}\partial_{z}\phi\right]\right) \nonumber \\
 &  & +g^{22}\left(\partial_{2}\phi^{*}-\frac{1}{2}{\rm i}m\rho\omega_{0}\left[\Theta_{3}\partial_{\rho}\phi^{*}-\Theta_{1}\partial_{z}\phi^{*}\right]\right)\left(\partial_{2}\phi+\frac{1}{2}{\rm i}m\rho\omega_{0}\left[\Theta_{3}\partial_{\rho}\phi-\Theta_{1}\partial_{z}\phi\right]\right) \nonumber \\
 &  & +g^{33}\left(\partial_{3}\phi^{*}-m\omega_{0}z\phi^{*}-\frac{1}{2}{\rm i}m\omega_{0}\left[-\Theta_{2}\partial_{\rho}\phi^{*}+\Theta_{1}\frac{1}{\rho}\partial_{\varphi}\phi^{*}\right]\right) \nonumber \\
 &  & \left.\times\left(\partial_{3}\phi+m\omega_{0}z\phi+\frac{1}{2}{\rm i}m\omega_{0}\left[-\Theta_{2}\partial_{\rho}\phi+\Theta_{1}\frac{1}{\rho}\partial_{\varphi}\phi\right]\right)+m^{2}\phi^{*}\phi\right]\,.
\end{eqnarray}
The corresponding Euler-Lagrange equation is: 
\begin{align}
0= & -m^{2}\phi+3m\omega_{0}\phi+\left(m\omega_{0}\rho\right)^{2}\phi+\left(m\omega_{0}z\right)^{2}\phi+\frac{1}{2}{\rm i}\Theta_{2}m^{2}\omega_{0}^{2}\frac{z}{\rho}\phi \nonumber \\
 & +2m\omega_{0}\rho\partial_{1}\phi+2m\omega_{0}z\partial_{3}\phi-{\rm i}m\omega_{0}\omega\Theta_{3}\partial_{0}\phi+{\rm i}m\omega_{0}\omega^{2}\Theta_{3}\partial_{2}\phi+{\rm i}m\omega_{0}\frac{1}{\rho}\Theta_{2}\partial_{3}\phi \nonumber \\
 & +\frac{1}{4}m^{2}\omega_{0}^{2}\left[\frac{\Theta_{2}^{2}}{\rho^{2}}-\left(3\omega^{2}-\frac{1}{\eta^{2}\rho^{2}}\right)\Theta_{3}^{2}\right]\rho\partial_{1}\phi+\frac{1}{4}m^{2}\omega_{0}^{2}\left(3\omega^{2}-\frac{1}{\eta^{2}\rho^{2}}\right)\left[\Theta_{3}\Theta_{1}\right]\rho\partial_{3}\phi \nonumber \\
 & -\partial_{0}^{2}\phi+2\omega\partial_{0}\partial_{2}\phi+\partial_{1}^{2}\phi+\frac{1}{\rho}\partial_{1}\phi+\left(\frac{1}{\eta^{2}\rho^{2}}-\omega^{2}\right)\partial_{2}^{2}\phi+\partial_{3}^{2}\phi \nonumber \\
 & -\left(\frac{1}{\eta^{2}\rho^{2}}-\omega^{2}\right)\left(\frac{1}{2}{\rm i}m\rho\omega_{0}\right)^{2}\left[\Theta_{3}^{2}\partial_{1}^{2}\phi-2\left(\Theta_{1}\Theta_{3}\right)\partial_{3}\partial_{1}\phi+\Theta_{1}^{2}\partial_{3}^{2}\phi\right] \nonumber \\
 & +\left(\frac{1}{2}{\rm i}m\omega_{0}\right)^{2}\left[-\Theta_{2}^{2}\partial_{1}^{2}\phi+2\Theta_{1}\Theta_{2}\frac{1}{\rho}\partial_{1}\partial_{2}\phi-\left(\Theta_{1}^{2}+\Theta_{3}^{2}\right)\frac{1}{\rho^{2}}\partial_{2}^{2}\phi+2\Theta_{2}\Theta_{3}\frac{1}{\rho}\partial_{3}\partial_{2}\phi-\Theta_{2}^{2}\partial_{3}^{2}\phi\right].
\end{align}
Comparison of this equation with the field equation containing the
contribution of the magnetic vector potential does not reveal a manifest
mapping from $\mathbf{A}=\left(A_{\rho},A_{\varphi},A_{z}\right)$
to $\mathbf{\Theta}=\left(\Theta_{\rho},\Theta_{\varphi},\Theta_{z}\right)$
such that a magnetic field could be described by non-commutativity
among space coordinates. This is the physical reason for our choice
to take into account only non-commutativity in the momentum space.


\subsection{Solution to the non-commutative-momentum-space field equation\label{subsec:Solution}}

In this subsection, we discuss the solutions of Eq.~(\ref{EMNC}).
Since we are interested in the time-independent solutions, we define
the field $\psi$ by 
\begin{equation}
\phi\left(\rho,\varphi,z,t\right)=e^{-i{\cal E}t}\psi\left(\rho,\varphi,z\right),
\end{equation}
As usual, ${\cal E}$ is interpreted as the energy and we perform
separation of variables in the space part: 
\begin{equation}
\psi\left(\rho,\varphi,z\right)=R\left(\rho\right)\Phi\left(\varphi\right)Z\left(z\right).\label{separation}
\end{equation}
By substituting this ansatz into Eq. (\ref{EMNC}), it becomes 
\begin{eqnarray}
2\left(\frac{1}{\eta^{2}\rho^{2}}-\omega^{2}\right)\left(i\frac{1}{2}\Omega_{3}\rho-i\frac{1}{2}\Omega_{1}z\right)\frac{\partial_{2}\varPhi\left(\varphi\right)}{\varPhi\left(\varphi\right)}-2i{\cal E}\omega\frac{\partial_{2}\varPhi\left(\varphi\right)}{\varPhi\left(\varphi\right)}+\left(\frac{1}{\eta^{2}\rho^{2}}-\omega^{2}\right)\frac{\partial_{2}^{2}\varPhi\left(\varphi\right)}{\varPhi\left(\varphi\right)}\nonumber \\
+2\left(m\omega_{0}z-\frac{1}{2}i\Omega_{2}\rho\right)\frac{\partial_{3}Z\left(z\right)}{Z\left(z\right)}+\frac{\partial_{3}^{2}Z\left(z\right)}{Z\left(z\right)}\nonumber \\
+\frac{\partial_{1}^{2}R\left(\rho\right)}{R\left(\rho\right)}+\frac{1}{\rho}\frac{\partial_{1}R\left(\rho\right)}{R\left(\rho\right)}+2\left(m\omega_{0}\rho+\frac{1}{2}i\Omega_{2}z\right)\frac{\partial_{1}R\left(\rho\right)}{R\left(\rho\right)}\nonumber \\
+{\cal E}^{2}-m^{2}+3m\omega_{0}+\left(m\omega_{0}\rho+\frac{1}{2}i\Omega_{2}z\right)^{2}+\left(m\omega_{0}z-\frac{1}{2}i\Omega_{2}\rho\right)^{2}\nonumber \\
+\left(\frac{1}{\eta^{2}\rho^{2}}-\omega^{2}\right)\left(\frac{1}{2}i\Omega_{3}\rho-\frac{1}{2}i\Omega_{1}z\right)^{2}+\frac{1}{2}i\Omega_{2}\frac{z}{\rho}-2i{\cal E}\omega\left(i\frac{1}{2}\Omega_{3}\rho-\frac{1}{2}i\Omega_{1}z\right)=0.\label{EM1}
\end{eqnarray}

The outcome of Eq. \eqref{separation} is not directly effective because
several terms involve the product of coordinates $\rho$ and $z$.
In order to make further progress, we choose 
\begin{equation}
\Omega_{1}=\Omega_{2}=0.\label{Om1Om2}
\end{equation}
This particular choice is not only convenient from the computational
standpoint, but also physically meaningful as discussed in Section
\ref{subsec:MagField} because of the mapping between the momentum-space
non-commutativity with $\Omega_{3}\neq0$ and a constant magnetic
field pointing in the $z$-direction \cite{Delduc}. The remaining
non-commutativity parameter couples to the frame angular velocity
$\omega$ but not to the oscillator frequency $\omega_{0}$.

The solution of Eq. (\ref{EM1}) is built by further imposing the
ansatz 
\begin{equation}
\Phi=e^{iL\varphi},\label{eq:AnsatzPhi}
\end{equation}
where 
\begin{equation}
L=0,\pm1,\pm2,\pm3,\cdots.\label{eq:QuantizationL}
\end{equation}
The quantization of $L$ stems from the familiar periodic boundary
condition upon the azimuthal function $\Phi\left(\varphi\right)$.
From Eq. (\ref{eq:AnsatzPhi}), we see that Eq. \eqref{EM1} decouples
into a $\rho$-dependent part, 
\begin{align}
\frac{\partial_{1}^{2}R\left(\rho\right)}{R\left(\rho\right)}+\frac{1}{\rho}\frac{\partial_{1}R\left(\rho\right)}{R\left(\rho\right)}+2m\omega_{0}\rho\frac{\partial_{1}R\left(\rho\right)}{R\left(\rho\right)}\nonumber \\
+{\cal E}^{2}-m^{2}+3m\omega_{0}+\left(m\omega_{0}\rho\right)^{2}-\left(\frac{1}{\eta^{2}\rho^{2}}-\omega^{2}\right)\left(\frac{1}{2}\Omega_{3}\rho\right)^{2}+{\cal E}\omega\Omega_{3}\rho\nonumber \\
-2L\left(\frac{1}{\eta^{2}\rho^{2}}-\omega^{2}\right)\frac{1}{2}\Omega_{3}\rho+2L{\cal E}\omega-L^{2}\left(\frac{1}{\eta^{2}\rho^{2}}-\omega^{2}\right) & =2km\label{eq:DiffEqR}
\end{align}
and a $z$-dependent sector, 
\begin{equation}
\frac{\partial_{3}^{2}Z\left(z\right)}{Z\left(z\right)}+2m\omega_{0}z\frac{\partial_{3}Z\left(z\right)}{Z\left(z\right)}+\left(m\omega_{0}z\right)^{2}=-2mk,\label{eq:DiffEqZ(z)}
\end{equation}
where $k$ is a separation constant with units of energy. The solution
of (\ref{eq:DiffEqZ(z)}) is 
\begin{equation}
Z\left(\zeta\right)=C_{1}e^{-\frac{\zeta^{2}}{2}-\zeta\sqrt{1-K}}+C_{2}e^{-\frac{\zeta^{2}}{2}+\zeta\sqrt{1-K}}\label{eq:Z(zeta)}
\end{equation}
with the definitions 
\begin{equation}
\zeta\equiv\sqrt{m\omega_{0}}z\qquad\text{and}\qquad K\equiv\frac{2k}{\omega_{0}}.\label{eq:zeta_K}
\end{equation}
Both $\zeta$ and $K$ are dimensionless constants.

The radial equation (\ref{eq:DiffEqR}) may be cast into the form
(organized in powers of $\rho$): 
\begin{align}
0= & \frac{\partial_{1}^{2}R\left(\rho\right)}{R\left(\rho\right)}+\frac{1}{\rho}\frac{\partial_{1}R\left(\rho\right)}{R\left(\rho\right)}+2m\omega_{0}\rho\frac{\partial_{1}R\left(\rho\right)}{R\left(\rho\right)} \nonumber \\
 & +\left(1+\epsilon^{2}\right)\left(m\omega_{0}\right)^{2}\rho^{2}+2\epsilon\left({\cal E}+L\omega\right)\left(m\omega_{0}\right)\rho \nonumber \\
 & +\left[\left({\cal E}+L\omega\right)^{2}-m^{2}+\left(3-K\right)m\omega_{0}-\left(\frac{1}{2}\frac{\Omega_{3}}{\eta}\right)^{2}\right] \nonumber \\
 & -2\left(\frac{1}{2}\frac{\Omega_{3}}{\eta}\right)\left(\frac{L}{\eta}\right)\frac{1}{\rho}-\left(\frac{L}{\eta}\right)^{2}\frac{1}{\rho^{2}},
\end{align}
where the new parameter $\epsilon$, 
\begin{equation}
\epsilon\equiv\left(\frac{1}{2}\frac{\Omega_{3}\omega}{m\omega_{0}}\right)\,,\label{eq:epsilon}
\end{equation}
contains the non-commutativity parameter $\Omega_{3}$. It is convenient
to perform a change of variables 
\begin{equation}
\xi\equiv S\rho\,,\label{eq:xi(S,rho)}
\end{equation}
where $S$ could assume two forms: either 
\begin{equation}
S=S\left(m,\omega_{0}\right)\equiv\sqrt{m\omega_{0}}\label{eq:S(m,omega_0)}
\end{equation}
or 
\begin{equation}
S=S\left(m,\omega_{0},\omega,\Omega_{3}\right)\equiv\sqrt{m\omega_{0}}\left(1+\epsilon^{2}\right)^{\frac{1}{4}}\,.\label{eq:S(epsilon)}
\end{equation}
The form of $S$ in Eq. (\ref{eq:xi(S,rho)}) resembles the one in
the definition of the coordinate $\zeta$, Eq. (\ref{eq:zeta_K}).
The functional form of $S$ in Eq. (\ref{eq:S(epsilon)}) includes
the non-commutativity parameter $\Omega_{3}$.

The solution of the radial equation by using Eq. (\ref{eq:S(epsilon)})
leads to interesting results, such as dissipative quantized energy,
but it also causes problems, which we will explore in Appendix A.

Henceforth, we will work with the choice in Eq. (\ref{eq:xi(S,rho)}).
In terms of $\xi$, the radial equation reads: 
\begin{align}
0= & \frac{\partial^{2}R\left(\xi\right)}{\partial\xi^{2}}+\frac{1}{\xi}\frac{\partial R\left(\xi\right)}{\partial\xi}+2\xi\frac{\partial R\left(\xi\right)}{\partial\xi}+\left(1+\epsilon^{2}\right)\xi^{2}R\left(\xi\right)+2\frac{\left({\cal E}+L\omega\right)}{\sqrt{m\omega_{0}}}\left(\epsilon\right)\xi R\left(\xi\right)\nonumber \\
 & +\frac{1}{\left(m\omega_{0}\right)}\left[\left({\cal E}+L\omega\right)^{2}-m^{2}+\left(3-K\right)m\omega_{0}-\left(\frac{1}{2}\frac{\Omega_{3}}{\eta}\right)^{2}\right]R\left(\xi\right)\nonumber \\
 & -\frac{2}{\sqrt{m\omega_{0}}}\left(\frac{1}{2}\frac{\Omega_{3}}{\eta}\right)\left(\frac{L}{\eta}\right)\frac{1}{\xi}R\left(\xi\right)-\left(\frac{L}{\eta}\right)^{2}\frac{1}{\xi^{2}}R\left(\xi\right).\label{eq:DiffEqR(xi,epsilon)}
\end{align}
At this point, we proceed to analyze two broad possibilities: the
commutative case and the non-commutative scenario.

\subsubsection{Commutative case }

In this subsection, we obtain the solution for the commutative case,
which will serve as our reference when taking the commutative limit.
By taking $\Omega_{3}=0$ (which implies $\epsilon=0$) in Eq. (\ref{eq:DiffEqR(xi,epsilon)}),
we obtain 
\begin{align}
0= & \frac{\partial^{2}R\left(\xi\right)}{\partial\xi^{2}}+\frac{1}{\xi}\frac{\partial R\left(\xi\right)}{\partial\xi}+2\xi\frac{\partial R\left(\xi\right)}{\partial\xi}+\xi^{2}R\left(\xi\right)\nonumber \\
 & +\frac{1}{\left(m\omega_{0}\right)}\left[\left({\cal E}+L\omega\right)^{2}-m^{2}+\left(3-K\right)m\omega_{0}\right]R\left(\xi\right)-\left(\frac{L}{\eta}\right)^{2}\frac{1}{\xi^{2}}R\left(\xi\right)\,.\label{eq:DiffEqR_commut}
\end{align}
The solution of this equation is in terms of the first-kind Bessel
function $J_{\alpha}(x)$: 
\begin{equation}
R\left(\xi\right)=C_{\text{c}}e^{-\frac{\xi^{2}}{2}}J_{\frac{L}{\eta}}\left(\alpha_{\text{c}}\xi\right)\,,\qquad\alpha_{\text{c}}\equiv\sqrt{\frac{-m^{2}+\left(1-K\right)m\omega_{0}+\left({\cal E}+L\omega\right)^{2}}{m\omega_{0}}}\,.\label{eq:R_commut}
\end{equation}
The quantity $C_{\text{c}}$ is an integration constant. In principle,
the differential equation also admits another solution in terms of
the Bessel $Y_{\frac{L}{\eta}}\left(\alpha_{\text{c}}\xi\right)$,
but this one diverges in the limit $\xi\rightarrow0$ and is thus
discarded.

The Bessel functions of the first kind $J_{\alpha}$ and of the second
kind $Y_{\alpha}$ are linearly independent solutions to the ordinary
second-order linear differential equation \cite{Arfken} 
\begin{equation}
x^{2}\frac{d^{2}y}{dx^{2}}+x\frac{dy}{dx}+\left(x^{2}-\alpha^{2}\right)y=0,
\end{equation}
called Bessel equation. This equation appears when solving the Laplace
equation and the Helmholtz equation in cylindrical or spherical coordinates
via separation of variables. Its applications span from quantum mechanics
to thermodynamics, hydrodynamics and acoustics. In fact, the Bessel
functions are common place as solutions of the Schr\"odinger equation
in spherical and cylindrical coordinates for the free particle; in
the description of heat conduction in cylindrical bars; in the description
of motion of floating bodies, and for modelling the vibration of thin
circular membranes, among others \cite{Kreyszig}. The series expansion
of $J_{\alpha}\left(x\right)$ around $x=0$ is found by the Frobenius
method and gives 
\begin{equation}
J_{\alpha}\left(x\right)=\sum_{m=0}^{\infty}\frac{\left(-1\right)^{m}}{m!\Gamma\left(m+\alpha+1\right)}\left(\frac{x}{2}\right)^{2m+\alpha},
\end{equation}
where $\Gamma\left(z\right)$ is the gamma function \cite{Arfken}.

As it happens, in quantum mechanics, for a particle confined in an
infinite spherical well \cite{Griffiths}, the energy quantization
is enforced via a boundary condition where the wave function should
be zero. In our context the boundary is set by the hard-wall condition
at $\rho=1/\omega\eta$. The roots of the Bessel functions do not
have an analytical closed form, but we will calculate them numerically
in Section \ref{roots}.


\subsubsection{Non-commutative case}

We will take a progressive approach, meaning that we shall seek solutions
of Eq. (\ref{eq:DiffEqR(xi,epsilon)}) with different orders of the
parameter $\epsilon$, defined in Eq. \eqref{eq:epsilon}, by keeping
terms up to order $\epsilon^{2}$ in the first place, then accounting
for orders of $\epsilon^{1}$, and, finally considering only terms
of $\epsilon^{0}$ in the radial differential equation. The reason
for this will become clear below.


\paragraph{$\epsilon^{2}$-order}

Here we keep all powers of $\epsilon$ in the differential equation
(\ref{eq:DiffEqR(xi,epsilon)}), which exhibits terms up to order
$\epsilon^{2}$. The solution --given by the Maple Software \cite{Maple}
-- is in terms of the biconfluent Heun solution \cite{Arriola}:
\begin{equation}
R\left(\xi\right)=C_{\text{nc}}\xi^{L/\eta}\exp\left\{ -\frac{\xi^{2}}{2}\left(1-i\epsilon\right)+i\frac{\left(\mathcal{E}+\omega L\right)}{\sqrt{m\omega_{0}}}\xi\right\} \text{HeunB}\left(\alpha,\beta,\gamma,\delta,\alpha_{\text{nc}}\xi\right)\,,\label{eq:R_epsilon2}
\end{equation}
where $C_{\text{nc}}$ is a constant, 
\begin{equation}
\begin{split} & \alpha\equiv2\frac{L}{\eta},\\
 & \beta\equiv i\frac{2^{3/2}\left(\Omega_{3}\omega\right)\left(m\omega_{0}\right)^{2}\left(\mathcal{E}+\omega L\right)}{\sqrt{m\omega_{0}}\left[-\left(\Omega_{3}\omega m\omega_{0}\right)^{2}\right]^{3/4}},\\
 & \gamma\equiv-2\frac{\left(\Omega_{3}\omega\right)^{2}\left(m\omega_{0}\right)^{3}\left[\left(m^{2}+m\omega_{0}\left(K-1\right)\right)\eta^{2}+\frac{\Omega_{3}^{2}}{4}\right]}{\eta^{2}\left[-\left(\Omega_{3}\omega m\omega_{0}\right)^{2}\right]^{3/2}},\\
 & \delta\equiv i\frac{2^{3/2}\left(\Omega_{3}L\right)\left(m\omega_{0}\right)}{\eta^{2}\sqrt{m\omega_{0}}\left[-\left(\Omega_{3}\omega m\omega_{0}\right)^{2}\right]^{1/4}},
\end{split}
\label{eq:Heun_parameters}
\end{equation}
and 
\begin{equation}
\alpha_{\text{nc}}=-i\frac{\sqrt{2}\left[-\left(\Omega_{3}\omega m\omega_{0}\right)^{2}\right]^{1/4}}{2m\omega_{0}}\,.\label{eq:alpha_nc}
\end{equation}

Heun's equation is the following second-order linear ordinary differential
equation \cite{Heun,NIST}: 
\begin{equation}
\frac{d^{2}w}{dz^{2}}+\left[\frac{\gamma}{z}+\frac{\delta}{z-1}+\frac{\varepsilon}{z-a}\right]\frac{dw}{dz}+\frac{\alpha\beta z-q}{z\left(z-1\right)\left(z-a\right)}w=0,
\end{equation}
where $\varepsilon=\alpha+\beta-\gamma-\delta+1$. The accessory parameter
$q$ is a complex number. In general, the above equation presents
four regular singular points, namely: 0, 1, $a$ and $\infty$. Its
importance is connected to the fact that every second-order ordinary
differential equation extended to the complex plane with four singular
points can be transformed to Heun's equation by a suitable change
of variables. This includes the Lam\'e equation (appearing when solving
Laplace equation in elliptic coordinates) and the hypergeometric equation
\cite{Arfken}. Hypergeometric functions are special functions represented
by a series and encompass many other special functions (such as Legendre
polynomials); the hypergeometric functions are general solutions of
second-order linear ordinary differential equations with three singular
points. Two or more regular singularities of Heun's equation may coalesce
into irregular singularities; this phenomenon gives rise to distinct
confluent types of Heun's equation. (An analogous process takes place
for the hypergeometric differential equation.) The biconfluent Heun
equation showcases two irregular (rank 2) singularities at 0 and $\infty$;
moreover, it is characterized by the differential equation \cite{GERG2019}
\begin{equation}
\frac{d^{2}w}{dz^{2}}+\left(\gamma+z\right)z\frac{dw}{dz}+\left(\alpha z-q\right)w=0.
\end{equation}
(Recall that $\alpha$, $\beta$, $\gamma$ and $\delta$ are related
through the parameter $\varepsilon$).

As a solution, Eq. (\ref{eq:R_epsilon2}) is problematic with respect
to the energy quantization. In fact, the quantization condition for
the biconfluent Heun function \cite{Arriola} is $\gamma-2-\alpha=2n$
$\left(n=1,2,3,\dots\right)$. It does not lead to the energy quantization
since $\gamma$ does not contain $\mathcal{E}$. At best, this could
be interpreted as a quantization condition on $K$ defined in Eq.(\ref{eq:zeta_K}).
Another attempt to quantize the energy would be to look for the roots
of the biconfluent Heun function. The problem with this is that the
argument, or variable, $\alpha_{\text{nc}}\xi$ of the Heun function
also does not depend on the energy. Thus, even the numerical analysis
does not help us to quantize $\mathcal{E}$.

In addition, the commutative limit is not well defined: in the limit
$\Omega_{3}\rightarrow0$, the variable $\alpha_{\text{nc}}\xi$ is
ill defined, and the parameters $\beta$ and $\gamma$ diverge --see
Eqs. (\ref{eq:Heun_parameters}) and (\ref{eq:alpha_nc}). We are
thus led to disregard this case.


\paragraph{$\epsilon^{1}$-order}

In the face of the difficulties with the $\epsilon^{2}$-order, we
try and keep the terms scaling with $\epsilon^{1}$ in Eq. (\ref{eq:DiffEqR(xi,epsilon)})
but neglect the term depending on $\epsilon^{2}$. This is justified
in the context where $\Omega_{3}\omega\ll m\omega_{0}$, i.e. the
combination of non-commutativity and frame rotation is subdominant
with respect to the rest mass and the oscillator's frequency. That
is the standpoint we adopt henceforth.

Accordingly, we set the term $\epsilon^{2}$ equals zero in Eq. (\ref{eq:DiffEqR(xi,epsilon)})
and input the resulting differential equation in a computer algebra
system to check for possible solutions. In this case, neither Maple
nor Mathematica can return a solution to the differential equation.
As an alternative, we resort to the Frobenius method for building
a solution. In this manner, it is possible to obtain a quantization
condition for the energy. However, we end up with the caveat of fixing
a large number of conditions on the coefficients of the series: at
least four coefficients have to be fixed, as discussed in Appendix
B. This is troublesome since we have a limited number of physical
parameters to deal with. Thus, we will abandon this possibility and
consider the even more simple case where terms scaling with $\epsilon$
and $\epsilon^{2}$ are neglected and only explicit term on the non-commutative
parameter $\Omega_{3}$ is kept in Eq. (\ref{eq:DiffEqR(xi,epsilon)}).


\paragraph{$\epsilon^{0}$-order}

We can rewrite the differential equation (\ref{eq:DiffEqR(xi,epsilon)})
as: 
\begin{align}
0= & \frac{\partial^{2}R\left(\xi\right)}{\partial\xi^{2}}+\frac{1}{\xi}\frac{\partial R\left(\xi\right)}{\partial\xi}+2\xi\frac{\partial R\left(\xi\right)}{\partial\xi}+\left(1+\epsilon^{2}\right)\xi^{2}R\left(\xi\right)\nonumber \\
 & +\frac{1}{\left(m\omega_{0}\right)}\left\{ \left({\cal E}+L\omega\right)^{2}\left[1+2\frac{1}{\left({\cal E}+L\omega\right)}\frac{m\omega_{0}}{\eta\omega}\epsilon\frac{\xi}{\xi_{0}}\right]-m^{2}+\left(3-K\right)m\omega_{0}-\left(\frac{1}{2}\frac{\Omega_{3}}{\eta}\right)^{2}\right\} R\left(\xi\right)\nonumber \\
 & -\frac{2}{\sqrt{m\omega_{0}}}\left(\frac{1}{2}\frac{\Omega_{3}}{\eta}\right)\left(\frac{L}{\eta}\right)\frac{1}{\xi}R\left(\xi\right)-\left(\frac{L}{\eta}\right)^{2}\frac{1}{\xi^{2}}R\left(\xi\right)\,,\label{eq:DiffEqR_epsilons}
\end{align}
where 
\begin{equation}
\xi_{0}=S\rho_{0}=\frac{\sqrt{m\omega_{0}}}{\eta\omega}\label{eq:xi_hw}
\end{equation}
is the hard-wall in the variable $\xi$ defined in terms of $\rho$
in Eq. \eqref{eq:xi(S,rho)}. In order to neglect the terms scaling
with $\epsilon^{2}$ and $\epsilon^{1}$ in Eq. (\ref{eq:DiffEqR_epsilons})
there are two conditions that must be matched. The first is to assume
$\Omega_{3}\omega\ll m\omega_{0}$; this enables us to discard the
term with $\epsilon^{2}$ in accordance with what was done in the
previous case. The second is to take 
\begin{equation}
2\frac{1}{\left({\cal E}+L\omega\right)}\frac{m\omega_{0}}{\eta\omega}\epsilon\frac{\xi}{\xi_{0}}\ll1.
\end{equation}
The physically allowed region is within the interval $0\leqslant\xi\leqslant\xi_{0}$,
which implies that, at most $\frac{\xi}{\xi_{0}}=1$. Accordingly,
the above condition is satisfied if $\epsilon$ is such that 
\begin{equation}
\epsilon\ll\frac{1}{2}\frac{\eta\omega}{m\omega_{0}}\left({\cal E}+L\omega\right)\Rightarrow\frac{\Omega_{3}}{\eta}\ll\left({\cal E}+L\omega\right).\label{eq:cond_epsilon}
\end{equation}
In the face of this condition, we should also consider the term $\left(\Omega_{3}/2\eta\right)^{2}$
as negligibly small when compared to $\left({\cal E}+L\omega\right)^{2}$.

By considering the aforementioned approximations in Eq. (\ref{eq:DiffEqR_epsilons}),
together with the definition of auxiliary parameters 
\begin{equation}
\begin{split} & B\equiv\frac{1}{\left(m\omega_{0}\right)}\left[\left({\cal E}+L\omega\right)^{2}-m^{2}+\left(3-K\right)m\omega_{0}\right],\\
 & C\equiv\frac{2}{\sqrt{m\omega_{0}}}\left(\frac{1}{2}\frac{\Omega_{3}}{\eta}\right)\left(\frac{L}{\eta}\right),
\end{split}
\label{eq:DiffEqF_parameters}
\end{equation}
and the ansatz 
\begin{equation}
R\left(\xi\right)=F\left(\xi\right)e^{\nicefrac{-\xi^{2}}{2}}\xi^{\nicefrac{-1}{2}}\,,\label{eq:R(F)}
\end{equation}
which is motivated by the asymptotic limits imposed on Eq. \eqref{eq:DiffEqR_epsilons},
then this differential equation is reduced to 
\begin{align}
0 & =F^{\prime\prime}\left(\xi\right)+\left\{ \left(B-2\right)-\frac{C}{\xi}+\frac{\frac{1}{4}-\left(\frac{L}{\eta}\right)^{2}}{\xi^{2}}\right\} F\left(\xi\right).\label{eq:DiffEqF}
\end{align}
We used the notation $F^{\prime\prime}\left(\xi\right)\equiv\frac{d^{2}F\left(\xi\right)}{d\xi^{2}}$
in Eq. (\ref{eq:DiffEqF}). We can see that its solution is given
in terms of Whittaker functions \cite{Arfken}: 
\begin{equation}
F\left(\xi\right)=C_{\text{nc}}\,M_{-\frac{C}{2\sqrt{2-B}},\frac{L}{\eta}}\left(2\sqrt{2-B}\xi\right)\,.\label{eq:F(WhittM)}
\end{equation}

The second independent solution allowed for $F\left(\xi\right)$ is
in terms of the Whittaker $U_{\kappa,\mu}(x)$ function. However,
it presents divergence issues and is not physically acceptable. It
should be emphasized that the quantum number $L$ has to be restricted
to positive values so as to avoid divergences in the form of Eq. (\ref{eq:F(WhittM)})
near $\xi=0$.

The Whittaker functions $M_{\kappa,\mu}(x)$ and $U_{\kappa,\mu}(x)$
are two linear independent solutions of Whittaker's equation, 
\begin{equation}
\frac{d^{2}w}{dz^{2}}+\left(-\frac{1}{4}+\frac{\kappa}{z}+\frac{1/4-\mu^{2}}{z^{2}}\right)w=0,
\end{equation}
which presents a singular points at 0 (regular) and at $\infty$ (irregular).
The solutions read 
\begin{align}
 & {\displaystyle M_{\kappa,\mu}\left(z\right)=\exp\left(-z/2\right)z^{\mu+\tfrac{1}{2}}M\left(\mu-\kappa+\tfrac{1}{2},1+2\mu,z\right)}, \nonumber \\
 & U_{\kappa,\mu}\left(z\right)=\exp\left(-z/2\right)z^{\mu+\tfrac{1}{2}}U\left(\mu-\kappa+\tfrac{1}{2},1+2\mu,z\right),
\end{align}
and are given in terms of Kummer's confluent hypergeometric function
(or confluent hypergeometric function of the first kind) \cite{Arfken}
\begin{equation}
M(a,b,z)=\sum_{n=0}^{\infty}\frac{\left(a\right)_{n}z^{n}}{\left(b\right)_{n}n!}={}_{1}F_{1}(a;b;z),
\end{equation}
with the Pochhammer symbol defined by
\begin{align}
 & \left(a\right)_{0}=1,\nonumber \\
 & \left(a\right)_{n}=a(a+1)(a+2)\dots(a+n-1)\,,\label{eq:Pochhammer}
\end{align}
and Tricomi's confluent hypergeometric function (or confluent hypergeometric
function of the second kind)
\begin{equation}
U(a,b,z)=\frac{\Gamma(1-b)}{\Gamma(a+1-b)}M(a,b,z)+\frac{\Gamma(b-1)}{\Gamma(a)}z^{1-b}M(a+1-b,2-b,z).
\end{equation}
Therein, the parameters $\left(a,b\right)$ are the ones appearing
in the (Kummer's form of the) confluent hypergeometric equation 
\begin{equation}
{\displaystyle z\frac{d^{2}w}{dz^{2}}+(b-z)\frac{dw}{dz}-aw=0.}
\end{equation}
The function $U$ has a singularity at zero, unlike Kummer's function
$M(a,b,z)$.




\section{Energy quantization \label{Energy}}

The energy eigenvalues are sensitive to the interplay between the
Gaussian function and the Whittaker function $M_{\kappa,\mu}(x)$
appearing in the functional form of the solution $R\left(\xi\right)$
in Eq. (\ref{eq:R(F)}). In fact, this is related to the location
of hard-wall condition $\xi_{0}$ in Eq. \eqref{eq:xi_hw}. The method
for analyzing energy quantization depends on whether $\xi_{0}$ lies
at a finite distance from the origin, or is at infinity. These two
possibilities will be explored in the next subsections: hard-wall
at infinity in Section \ref{xi0infinity} and finite hard-wall in
Section \ref{roots}.


\subsection{Energy for the hard-wall in the limit $\xi_{0}\rightarrow\infty$:
Asymptotic limits\label{xi0infinity}}

If the hard-wall is very far from the origin, the behavior of the
radial solution will depend on the asymptotic limits of the Whittaker
function $M_{\kappa,\mu}(x)$. The functional form of these limits
are determined by the nature of the variable $x=2\sqrt{2-B}\xi$,
which could be real-valued or complex-valued.


\subsubsection{Complex-valued argument for the Whittaker function $M_{\kappa,\mu}(x)$\label{subsec:Complex-valued-argument}}

The variable $x$ will be a pure complex number if $2-B<0$, so that
Eq. \eqref{eq:DiffEqF_parameters} implies 
\begin{equation}
2-\frac{1}{\left(m\omega_{0}\right)}\left[\left({\cal E}+L\omega\right)^{2}-m^{2}+\left(3-K\right)m\omega_{0}\right]<0.
\end{equation}
In this case, $x=iy$, where $y$ is real, and the asymptotic behavior
for $M_{\kappa,\mu}(x)$ is \cite{NIST}: 
\begin{align}
M_{\kappa,\mu}(iy) & \sim\frac{\Gamma\left(1+2\mu\right)}{\Gamma\left(\frac{1}{2}+\mu-\kappa\right)}e^{\frac{iy}{2}}\left(iy\right)^{-\kappa}\sum_{s=0}^{\infty}\frac{\left(\frac{1}{2}-\mu+\kappa\right)_{s}\left(\frac{1}{2}+\mu+\kappa\right)_{s}}{s!}\left(iy\right)^{-s} \nonumber \\
 & +\frac{\Gamma\left(1+2\mu\right)}{\Gamma\left(\frac{1}{2}+\mu-\kappa\right)}e^{-\frac{iy}{2}\pm\left(\frac{1}{2}+\mu-\kappa\right)\pi i}\left(iy\right)^{\kappa}\sum_{s=0}^{\infty}\frac{\left(\frac{1}{2}+\mu-\kappa\right)_{s}\left(\frac{1}{2}-\mu-\kappa\right)_{s}}{s!}\left(-iy\right)^{-s}
\end{align}
where $\left(a\right)_{s}$ is the Pochhammer symbol of Eq. (\ref{eq:Pochhammer}).

Then, $M_{\kappa,\mu}(x)$ will assume a polynomial form under the
condition \cite{Arfken} 
\begin{equation}
\mu-\kappa+\frac{1}{2}=-n\qquad\left(n=0,1,2,3,...\right).
\end{equation}
This also leads to the quantization of energy. In fact, if we substitute
the values of the constants $B$ and $C$, from Eq. \eqref{eq:DiffEqF_parameters},
in the above condition, we obtain the energy eigenvalues:

\begin{equation}
{\cal E}_{n}=-L\omega\pm\sqrt{m^{2}+\left(K-1\right)m\omega_{0}-\frac{\left(\frac{1}{2}\frac{\Omega_{3}}{\eta}\right)^{2}\left(\frac{L}{\eta}\right)^{2}}{\left(\frac{1}{2}+\frac{L}{\eta}+n\right)^{2}}}.\label{eq:E_WhittM_complex_variable}
\end{equation}

The energy eigenvalues scale as $1/n$. It is of particular interest
that the energy quantization is lost either in the commutative case,
when $\Omega_{3}=0$, or when the angular momentum quantum number
is null, $L=0$. We observe that the contribution of the non-commutative
parameter is to decrease the modulus of the energy of the system.

Eq. \eqref{eq:E_WhittM_complex_variable} also shows that the angular
momentum quantum number couples with both the non-commutative parameter
and the angular velocity of the rotating frame. It is interesting
to note that the quantization is lost in the limit as $L/\eta\gg n$
when the contribution of the non-commutative parameter freezes out.

The sign $\pm$ in Eq. (\ref{eq:E_WhittM_complex_variable}) could
be interpreted as corresponding to the energy spectrum of particle
(plus sign) and anti-particles (negative sign). This interpretation
can be applied to other similar expressions later in this paper.

The energy ${\cal E}_{n}$ increases as the mass $m$ increases, as
expected. In particular, the large-mass regime leads to the the non-relativistic
limit 
\begin{align}
{\cal E}_{n} & \approx-L\omega\pm m\pm\frac{1}{2}\left(K-1\right)\omega_{0}\mp\frac{1}{2}\frac{1}{m}\frac{\left(\frac{1}{2}\frac{\Omega_{3}}{\eta}\right)^{2}\left(\frac{L}{\eta}\right)^{2}}{\left(\frac{1}{2}+\frac{L}{\eta}+n\right)^{2}}.\label{eq:E_WhittM_complex_variable_non-rel}
\end{align}
Eq. (\ref{eq:E_WhittM_complex_variable_non-rel}) also reveals that
the quantization is lost when $m\rightarrow\infty$, and that in this
limit, the non-commutativity is irrelevant.


\subsubsection{Real-valued argument for $M_{\kappa,\mu}(x)$ \label{subsec:Real-valued-argument}}

Unlike the Section \ref{subsec:Complex-valued-argument}, here we
assume $2-B>0$, and Eq. \eqref{eq:DiffEqF_parameters} leads to 
\begin{equation}
2-\frac{1}{\left(m\omega_{0}\right)}\left[\left({\cal E}+L\omega\right)^{2}-m^{2}+\left(3-K\right)m\omega_{0}\right]>0.
\end{equation}
Then the argument of $M_{\kappa,\mu}(x)$ is real and its asymptotic
behavior is: 
\begin{equation}
M_{\kappa,\mu}(x)\sim\frac{\Gamma\left(1+2\mu\right)}{\Gamma\left(\frac{1}{2}+\mu-\kappa\right)}e^{\frac{x}{2}}x^{-\kappa}\sum_{s=0}^{\infty}\frac{\left(\frac{1}{2}-\mu+\kappa\right)_{s}\left(\frac{1}{2}+\mu+\kappa\right)_{s}}{s!}x^{-s},\qquad\mu-\kappa\neq-\frac{1}{2},-\frac{3}{2},\ldots
\end{equation}

By considering the dominant term in the series above, the radial solution
becomes 
\begin{equation}
R\left(\xi\right)\sim C_{1}\xi^{-\left(\kappa+\frac{1}{2}\right)}e^{-\frac{1}{2}\left(\xi-\xi_{m}\right)^{2}}
\end{equation}
where 
\begin{equation}
\xi_{m}=\sqrt{\frac{m^{2}-\left({\cal E}+L\omega\right)^{2}+\left(K-1\right)m\omega_{0}}{m\omega_{0}}}.
\end{equation}
We have expressed the radial function as the product of a polynomial
by a Gaussian function with standard deviation $\sigma=1$ and mean
$\xi_{m}$.

Now let us impose that the wave function be zero at the hard-wall
position. We know that the Gaussian does not vanish at any point in
its domain. However, it quickly tends to zero as we move away from
$\xi_{m}$. If the hard-wall is larger than $k_{G}\sigma$, where
$\sigma$ denotes the standard deviation and $k_{G}$ is a real number
sufficiently large (e.g. $k_{G}\gtrsim6$), then the radial function
can be considered small enough so that its contribution to the probability
distribution is negligible near the hard-wall. If this condition is
satisfied, then the quantization condition is completely analogous
to the case discussed in Section \ref{subsec:Complex-valued-argument}
with the energy eigenvalues given precisely by Eq. (\ref{eq:E_WhittM_complex_variable}).
Otherwise, if the condition $\left(\xi_{0}-\xi_{m}\right)>k_{G}\sigma$
is not satisfied, then we have to enforce the boundary condition $R\left(\xi_{0}\right)=0$,
and a quantization condition can only be obtained by numerical calculations.
This is the case when the hard-wall is located in a definite finite
value of $\xi$. We shall analyze this possibility in Section \ref{roots}.


\subsection{Energy for a finite hard-wall: Numerical results\label{roots}}

This section deals with the energy eigenvalues admissible by the wave
function for a finite hard-wall. The roots of the special functions
entering the radial part of the wave function do not exhibit a closed
form in terms of the physical parameters and an integer number. This
is the reason why we need to utilize a numerical method to compute
the roots of the wave function at the boundary for selected values
of the free parameters in our model. We will perform the computations
in the commutative and non-commutative instances.

The commutative case admits a radial function as given in Eq. (\ref{eq:R_commut}).
The quantization of the energy is obtained when $r_{\text{c}}=\alpha_{\text{c}}\xi_{0}$
is a root of the Bessel function. If we evaluate the first $n$ roots
$r_{\text{c},n}$, we can find the corresponding energies ${\cal E}_{n}:$
\begin{equation}
{\cal E}_{n_{\pm}}=-L\omega\pm\sqrt{m^{2}+\left(K-1\right)m\omega_{0}+\left(\eta\omega\,r_{\text{c},n}\right)^{2}}\,,\qquad n=1,2,3,\dots\label{eq:E_commut}
\end{equation}

The non-commutative case is described by the radial function in Eqs.
(\ref{eq:R(F)}) and (\ref{eq:F(WhittM)}). It is thus given in terms
of a Whittaker function, whose zeros at the hard-wall location determine
the energy eigenvalues. The roots are $r_{\text{nc}}=r_{\text{nc}}\left(\Omega_{3}\right)=2\sqrt{2-B}\xi_{0}$
and the corresponding energies ${\cal E}_{n}$ are: 
\begin{equation}
{\cal E}_{n_{\pm}}=-L\omega\pm\sqrt{m^{2}+\left(K-1\right)m\omega_{0}-\frac{1}{4}\left(\eta\omega\,r_{\text{nc},n}\right)^{2}}\,,\qquad n=1,2,3,\dots\label{eq:E_non-commut}
\end{equation}
Notice that the roots $r_{\text{nc},n}$ are a function of the non-commutative
parameter $\Omega_{3}$.

\begin{figure}[H]
\begin{centering}
\includegraphics[scale=0.4]{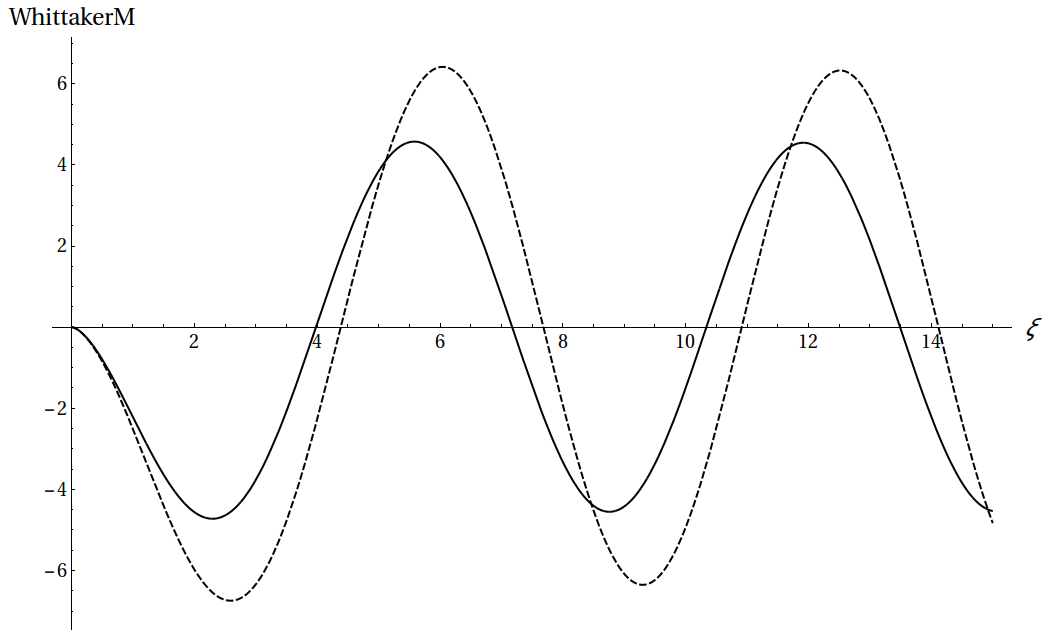} 
\par\end{centering}
\caption{Plot of the Whittaker function ${\normalcolor M_{\kappa,\mu}(x)}$
for $\Omega_{3}=0$ (continuous line) and $\Omega_{3}=0.1$ (dashed
line) for fixed values of the parameters $m=1$, $\omega_{0}=0.1$,
$\omega=0.5$, $\eta=0.9$, $L=1$ and $\mathcal{E}=0.5$.}

\label{Fig-1} 
\end{figure}

\begin{figure}[H]
\begin{centering}
\includegraphics[scale=0.5]{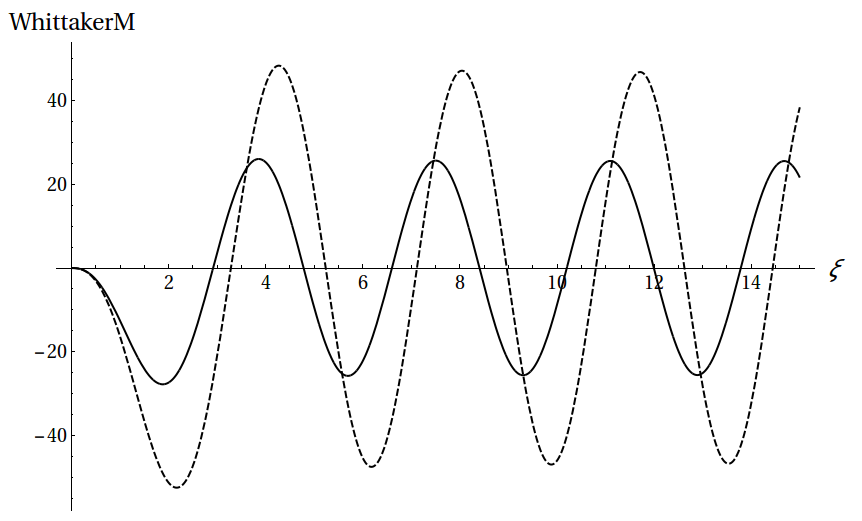} 
\par\end{centering}
\caption{Plot of the Whittaker function ${\normalcolor M_{\kappa,\mu}(x)}$
for $\Omega_{3}=0$ (continuous line) and $\Omega_{3}=0.1$ (dashed
line) for fixed values of the parameters $m=1$, $\omega_{0}=0.1$,
$\omega=1.0$, $\eta=0.5$, $L=1$ and $\mathcal{E}=0.1$.}

\label{Fig-2} 
\end{figure}

The presence of $\Omega_{3}$ displaces the zeros of the wave function
as can be seen in Figs. \ref{Fig-1} and \ref{Fig-2}.

The displacement of the roots of the wave function due to non-commutativity
implies that different values of the energy have to be chosen for
the commutative and non-commutative cases, cf. Eqs. (\ref{eq:E_commut})
and (\ref{eq:E_non-commut}). We use Mathematica \cite{Mathematica}
to determine the energy eigenvalues for both cases. These values are
given in Tables \ref{Tab-1} and \ref{Tab-2} for particular values
of the physical parameters $\left\{ m,\omega_{0},\omega,\eta,\Omega_{3}\right\} $,
of the quantum numbers $\left\{ n,L\right\} $, and of $K$.

\begin{table}[H]
\begin{centering}
\begin{tabular}{|c|c|c|c|c|c|}
\cline{3-6} 
\multicolumn{1}{c}{} &  & \multicolumn{4}{c|}{$L=0$}\tabularnewline
\cline{3-6} 
\multicolumn{1}{c}{} &  & \multicolumn{2}{c|}{$\Omega_{3}=0$} & \multicolumn{2}{c|}{$\Omega_{3}=0.1$}\tabularnewline
\hline 
\multirow{2}{*}{$\eta$} & \multirow{2}{*}{$n$} & \multicolumn{2}{c|}{$\left|\mathcal{E}_{\pm}\right|$} & \multicolumn{2}{c|}{$\left|\mathcal{E}_{\pm}\right|$}\tabularnewline
\cline{3-6} 
 &  & $K=0$  & $K=2$  & $K=0$  & $K=2$ \tabularnewline
\hline 
\multirow{3}{*}{0.9} & $1$  & 1.43913  & 1.50702  & 1.43913  & 1.50702 \tabularnewline
\cline{2-6} 
 & $2$  & 2.65903  & 2.69637  & 2.65903  & 2.69637 \tabularnewline
\cline{2-6} 
 & $3$  & 4.00807  & 4.03294  & 4.00807  & 4.03294 \tabularnewline
\hline 
\multirow{3}{*}{0.5 } & $1$  & 1.12314  & 1.2089  & 1.12314  & 1.2089 \tabularnewline
\cline{2-6} 
 & $2$  & 1.67465  & 1.73334  & 1.67465  & 1.73334 \tabularnewline
\cline{2-6} 
 & $3$  & 2.3623  & 2.40425  & 2.3623  & 2.40425 \tabularnewline
\hline 
\multirow{3}{*}{0.1} & $1$  & 0.956273  & 1.05568  & 0.956273  & 1.05568 \tabularnewline
\cline{2-6} 
 & $2$  & 0.988017  & 1.08452  & 0.988017  & 1.08452 \tabularnewline
\cline{2-6} 
 & $3$  & 1.0427  & 1.13456  & 1.0427  & 1.13456 \tabularnewline
\hline 
\end{tabular}
\par\end{centering}
\bigskip{}

\resizebox{\textwidth}{!}{

\begin{tabular}{|c|c|c|c|c|c|c|c|c|c|}
\cline{3-10} 
\multicolumn{1}{c}{} &  & \multicolumn{8}{c|}{$L=1$}\tabularnewline
\cline{3-10} 
\multicolumn{1}{c}{} &  & \multicolumn{4}{c|}{$\Omega_{3}=0$} & \multicolumn{4}{c|}{$\Omega_{3}=0.1$}\tabularnewline
\hline 
\multirow{2}{*}{$\eta$} & \multirow{2}{*}{$n$} & \multicolumn{2}{c|}{$\mathcal{E}_{+}$} & \multicolumn{2}{c|}{$\mathcal{E}_{-}$} & \multicolumn{2}{c|}{$\mathcal{E}_{+}$} & \multicolumn{2}{c|}{$\mathcal{E}_{-}$}\tabularnewline
\cline{3-10} 
 &  & $K=0$  & $K=2$  & $K=0$  & $K=2$  & $K=0$  & $K=2$  & $K=0$  & $K=2$\tabularnewline
\hline 
\multirow{3}{*}{0.9} & $1$  & 1.52706  & 1.57581  & $-$2.52706  & $-$2.57581  & 1.55421  & 1.60233  & $-$2.55421  & $-$2.60233\tabularnewline
\cline{2-10} 
 & $2$  & 2.86526  & 2.89484  & $-$3.86526  & $-$3.89484  & 2.88623  & 2.91564  & $-$3.88623  & $-$3.91564\tabularnewline
\cline{2-10} 
 & $3$  & 4.24752  & 4.26854  & $-$5.24752  & $-$5.26854  & 4.26449  & 4.28543  & $-$5.26449  & $-$5.28543\tabularnewline
\hline 
\multirow{3}{*}{0.5 } & $1$  & 1.09638  & 1.15783  & $-$2.09638  & $-$2.15783  & 1.14917  & 1.20873  & $-$2.14917  & $-$2.20873\tabularnewline
\cline{2-10} 
 & $2$  & 1.80827  & 1.8512  & $-$2.80827  & $-$2.8512  & 1.85459  & 1.89669  & $-$2.85459  & $-$2.89669\tabularnewline
\cline{2-10} 
 & $3$  & 2.55594  & 2.58849  & $-$3.55594  & $-$3.58849  & 2.59602  & 2.62815  & $-$3.59602  & $-$3.62815\tabularnewline
\hline 
\multirow{3}{*}{0.1} & $1$  & 0.693252  & 0.774304  & $-$1.69325  & $-$1.774304  & 0.448683  & 0.548809  & $-$1.44868  & $-$1.548809\tabularnewline
\cline{2-10} 
 & $2$  & 0.82268  & 0.896238  & $-$1.82268  & $-$1.896238  & 0.929379  & 0.997706  & $-$1.92938  & $-$1.997706\tabularnewline
\cline{2-10} 
 & $3$  & 0.954364  & 1.02157  & $-$1.95436  & $-$2.02157  & 1.06786  & 1.13039  & $-$2.06786  & $-$2.13039\tabularnewline
\hline 
\end{tabular}

}\caption{Values for the energies $\mathcal{E}_{nL}$ of the Whittaker function
$M_{\kappa,\mu}(x)=0$ with the fixed values $m=1,$ $\omega_{0}=0.1$
and $\omega=0.5$.}
\label{Tab-1} 
\end{table}

\begin{table}[H]
\begin{centering}
\begin{tabular}{|c|c|c|c|c|c|}
\cline{3-6} 
\multicolumn{1}{c}{} &  & \multicolumn{4}{c|}{$L=0$}\tabularnewline
\cline{3-6} 
\multicolumn{1}{c}{} &  & \multicolumn{2}{c|}{$\Omega_{3}=0$} & \multicolumn{2}{c|}{$\Omega_{3}=0.1$}\tabularnewline
\hline 
\multirow{2}{*}{$\eta$} & \multirow{2}{*}{$n$} & \multicolumn{2}{c|}{$\left|\mathcal{E}_{\pm}\right|$} & \multicolumn{2}{c|}{$\left|\mathcal{E}_{\pm}\right|$}\tabularnewline
\cline{3-6} 
 &  & $K=0$  & $K=2$  & $K=0$  & $K=2$ \tabularnewline
\hline 
\multirow{3}{*}{0.9} & $1$  & 2.36313  & 2.40507  & 2.36313  & 2.40507 \tabularnewline
\cline{2-6} 
 & $2$  & 5.05784  & 5.07757  & 5.05784  & 5.07757 \tabularnewline
\cline{2-6} 
 & $3$  & 7.84592  & 7.85866  & 7.84592  & 7.85866 \tabularnewline
\hline 
\multirow{3}{*}{0.5} & $1$  & 1.5316  & 1.59556  & 1.5316  & 1.59556 \tabularnewline
\cline{2-6} 
 & $2$  & 2.91853  & 2.95259  & 2.91853  & 2.95259 \tabularnewline
\cline{2-6} 
 & $3$  & 4.42964  & 4.45216  & 4.42964  & 4.45216 \tabularnewline
\hline 
\multirow{3}{*}{0.1} & $1$  & 0.978689  & 1.07603  & 0.978689  & 1.07603 \tabularnewline
\cline{2-6} 
 & $2$  & 1.09759  & 1.18521  & 1.09759  & 1.18521 \tabularnewline
\cline{2-6} 
 & $3$  & 1.28408  & 1.35973  & 1.28408  & 1.35973 \tabularnewline
\hline 
\end{tabular}
\par\end{centering}
\bigskip{}
\resizebox{\textwidth}{!}{

\begin{tabular}{|c|c|c|c|c|c|c|c|c|c|}
\cline{3-10} 
\multicolumn{1}{c}{} &  & \multicolumn{8}{c|}{$L=1$}\tabularnewline
\cline{3-10} 
\multicolumn{1}{c}{} &  & \multicolumn{4}{c|}{$\Omega_{3}=0$} & \multicolumn{4}{c|}{$\Omega_{3}=0.1$}\tabularnewline
\hline 
\multirow{2}{*}{$\eta$} & \multirow{2}{*}{$n$} & \multicolumn{2}{c|}{$\mathcal{E}_{+}$} & \multicolumn{2}{c|}{$\mathcal{E}_{-}$} & \multicolumn{2}{c|}{$\mathcal{E}_{+}$} & \multicolumn{2}{c|}{$\mathcal{E}_{-}$}\tabularnewline
\cline{3-10} 
 &  & $K=0$  & $K=2$  & $K=0$  & $K=2$  & $K=0$  & $K=2$  & $K=0$  & $K=2$\tabularnewline
\hline 
\multirow{3}{*}{0.9} & $1$  & 2.7062  & 2.73308  & $-$4.7062  & $-$4.73308  & 2.73601  & 2.76268  & $-$4.73601  & $-$4.76268\tabularnewline
\cline{2-10} 
 & $2$  & 5.52685  & 5.54216  & $-$7.52685  & $-$7.54216  & 5.54853  & 5.56378  & $-$7.54853  & $-$7.56378\tabularnewline
\cline{2-10} 
 & $3$  & 8.35179  & 8.36247  & $-$10.3518  & $-$10.3625  & 8.36903  & 8.3797  & $-$10.369  & $-$10.3797\tabularnewline
\hline 
\multirow{3}{*}{0.5} & $1$  & 1.73745  & 1.77374  & $-$3.73745  & $-$3.77374  & 1.79951  & 1.83501  & $-$3.79951  & $-$3.83501\tabularnewline
\cline{2-10} 
 & $2$  & 3.31422  & 3.33734  & $-$5.31422  & $-$5.33734  & 3.36409  & 3.38694  & $-$5.36409  & $-$5.38694\tabularnewline
\cline{2-10} 
 & $3$  & 4.88686  & 4.90383  & $-$6.88686  & $-$6.90383  & 4.92865  & 4.94549  & $-$6.92865  & $-$6.94549\tabularnewline
\hline 
\multirow{3}{*}{0.1} & $1$  & 0.730723  & 0.787569  & $-$2.73072  & $-$2.78757  & 2.3253  & 3.26675  & $-$4.3253  & $-$5.26675\tabularnewline
\cline{2-10} 
 & $2$  & 1.07314  & 1.12083  & $-$3.07314  & $-$3.12083  & 1.06036  & 1.10834  & $-$3.06036  & $-$3.10834\tabularnewline
\cline{2-10} 
 & $3$  & 1.40014  & 1.44145  & $-$3.40014  & $-$3.44145  & 1.39413  & 1.43554  & $-$3.39413  & $-$3.43554\tabularnewline
\hline 
\end{tabular}

}\caption{Values for the energies $\mathcal{E}_{nL}$ of the Whittaker function
$M_{\kappa,\mu}(x)=0$ with the fixed values $m=1,$ $\omega_{0}=0.1$
and $\omega=1$.}
\label{Tab-2} 
\end{table}

Tables \ref{Tab-1} and \ref{Tab-2} show that, for $L=0$, non-commutativity
has no effect on the energy eigenvalues, because of the coupling between
$\Omega_{3}$ and $L$ in the second-to-last term in Eq. (\ref{eq:DiffEqR_epsilons}).
For $L\neq0$, non-commutativity comes into effect such that,
as a general trend, the absolute value of $\mathcal{E}_{nL}$ increases
in the presence of non-commutativity. We also observe that the absolute
value of the energy is larger for a larger angular velocity $\omega$,
and mainly for a larger cosmic string parameter $\eta$.

The quantum number $n$ has the effect of increasing the energies
${\mathcal{E}}_{\pm}$ in absolute values. Although the majority of
the values of ${\mathcal{E}}_{\pm}$ increases with the increase of
$L$, a few exceptions in Tables \ref{Tab-1} and \ref{Tab-2} show
that there is no global pattern for the energy related to the quantum
number $L$.

Fig. \ref{Fig-E(eta,n,omega)} helps to visualize the effects mentioned
in the last two paragraphs. 

\begin{figure}[H]
\begin{centering}
\includegraphics[scale=0.33]{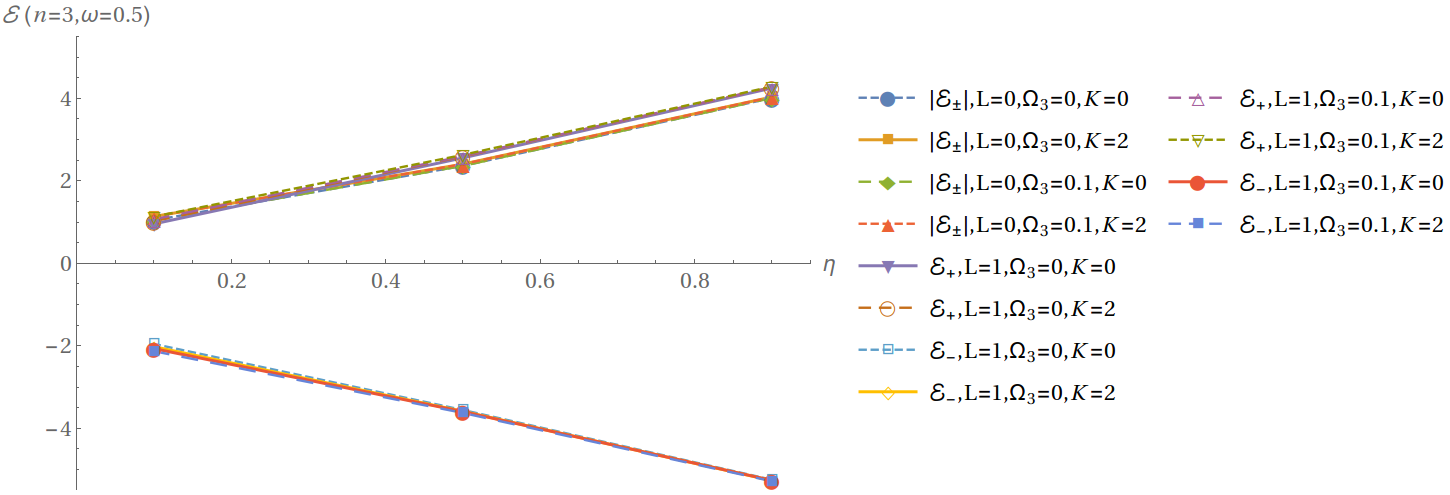}(a)$\quad$
\par\end{centering}
\begin{centering}
\includegraphics[scale=0.33]{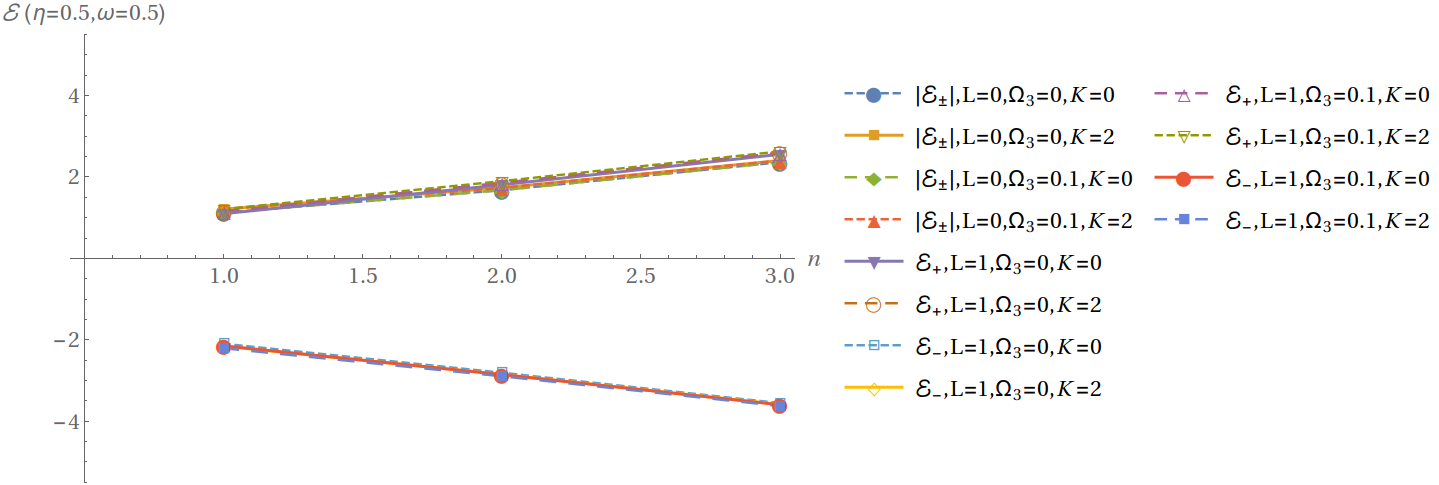}(b)$\quad$
\par\end{centering}
\begin{centering}
\includegraphics[scale=0.35]{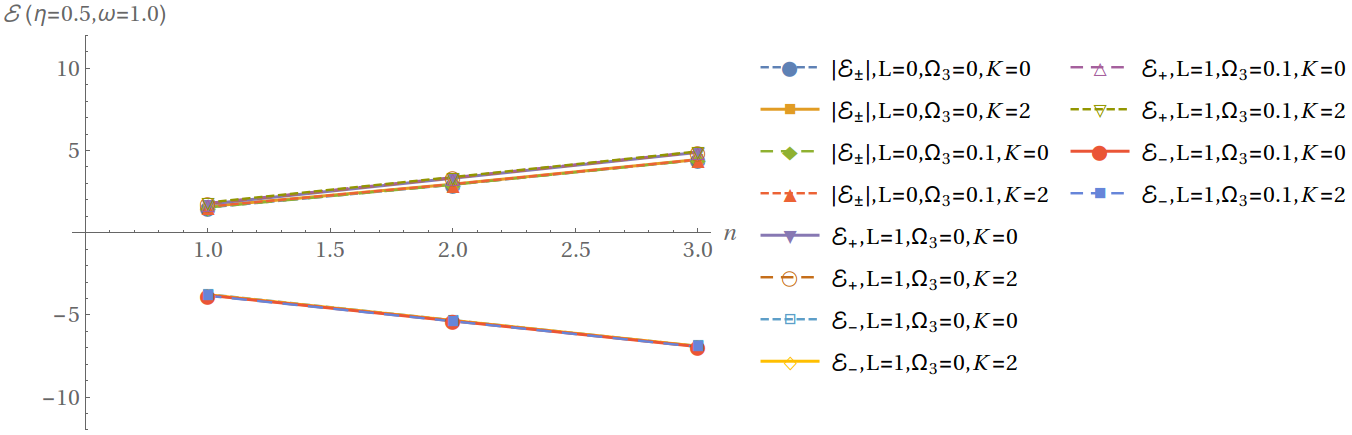}(c)$\quad$
\par\end{centering}
\caption{{Effects of the parameters $\left\{ \eta,n,\omega\right\} $ on the
energy eigenvalues ${\mathcal{E}}_{\pm}$. (a) Plot of $\mathcal{E}_{\pm}\left(\eta\right)$
for $n=3$ and $\omega=0.5$; (b) Plot of $\mathcal{E}_{\pm}\left(n\right)$
for $\eta=0.5$ and $\omega=0.5$; and (c) Plot of $\mathcal{E}_{\pm}\left(n\right)$
for $\eta=0.5$ and $\omega=1.0$. The absolute value of the energy
energy increases for larger values of $\eta$, from part (a); $\left|\mathcal{E}_{\pm}\right|$
increases as $n$ increases, from part (b); and the comparison of
parts (b) and (c) shows that $\left|\mathcal{E}_{\pm}\right|$ grows
with increasing $\omega$. The several lines in each part correspond
to different values of the parameters $L=0,1$, $\Omega_{3}=0.0,0.1$
and $K=0,2$.}}
\label{Fig-E(eta,n,omega)}
\end{figure}

In Fig. \ref{Fig-E(eta,n,omega)}(a) we set $n=3$ (the third energy
level), take $\omega=0.5$ (which selects the results in Table \ref{Tab-1})
and take $\eta=0.1,0.5,0.9$. The plot clearly shows that $\left|\mathcal{E}_{\pm}\right|$
increases with increasing values of $\eta$. Fig. \ref{Fig-E(eta,n,omega)}(b)
assumes $\omega=0.5$ (as in Table \ref{Tab-1} again), string parameter
$\eta=0.5$, and we let the principal quantum number $n=1,2,3$. The
graph shows that the values of $\left|\mathcal{E}_{\pm}\right|$ increase
as $n$ scales up. Furthermore, we can comclude that $\left|\mathcal{E}_{\pm}\right|$
increases as $\omega$ increases by comparing the values on the $y$-axis
of parts (b) and (c) of Fig. \ref{Fig-E(eta,n,omega)}, which display
$\mathcal{E}_{\pm}\left(n,\eta=0.5\right)$ for $\omega=0.5$ and
1.0, respectively.

\begin{figure}[H]
\begin{centering}
\includegraphics[scale=0.5]{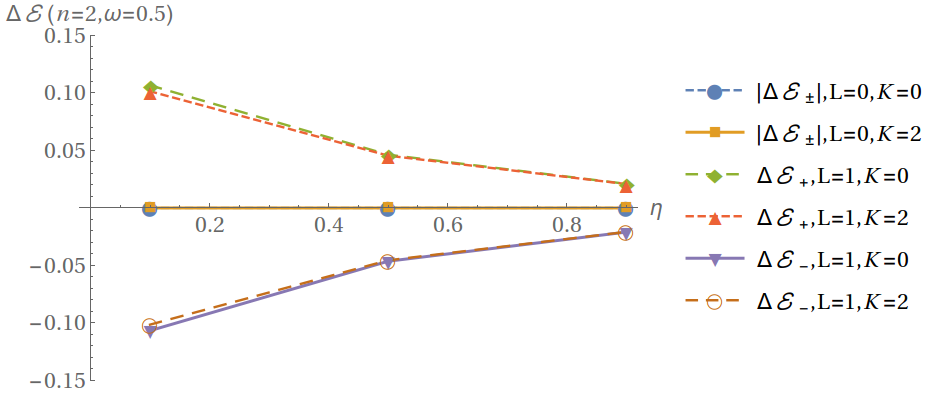}(a)$\quad$
\par\end{centering}
\begin{centering}
\includegraphics[scale=0.5]{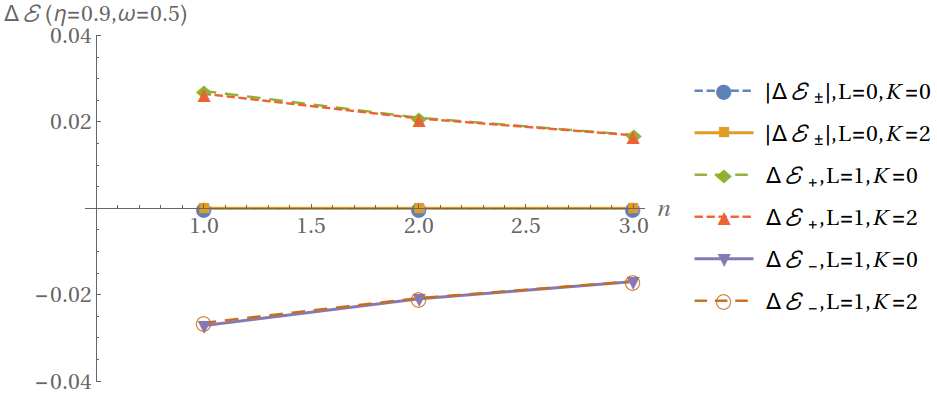}(b)$\quad$
\par\end{centering}
\caption{{Effect of the non-commutativity parameter $\Omega_{3}$ on the energy
eigenvalues $\mathcal{E}_{\pm}$. (a) Plot of $\Delta\mathcal{E}_{\pm}\equiv\mathcal{E}_{\pm}\left(\Omega_{3}=0.1\right)-\mathcal{E}_{\pm}\left(\Omega_{3}=0\right)$
as a function of the string parameter $\eta$ for a fixed energy level
$n=2$. (b) Plot of $\Delta\mathcal{E}_{\pm}$ as a function of the
principal quantum number $n$ for a fixed string parameter $\eta=0.9$.
Both parts (a) and (b) show that the effect of non-commutativity decreases
for increasing values of $\eta$ and $n$. The various lines in each
part correspond to different values of the parameters $\omega=0.5,1.0$,
$L=0,1$, and $K=0,2$.}}

\label{Fig-DeltaE(Omega3)}
\end{figure}

The effect of non-commutativity is displayed in Fig. \ref{Fig-DeltaE(Omega3)}.

We define the energy difference $\Delta\mathcal{E}_{\pm}\equiv\mathcal{E}_{\pm}\left(\Omega_{3}=0.1\right)-\mathcal{E}_{\pm}\left(\Omega_{3}=0\right)$
which vanishes if the non-commutativity parameter $\Omega_{3}$ has
no impact on the energy levels. We evaluate $\Delta\mathcal{E}_{\pm}$
as a function of $\eta$ (for a constant value of the principal quantum
number $n$) in Fig. \ref{Fig-DeltaE(Omega3)}(a): it shows that the
effect of non-commutativity on the values of $\left|\mathcal{E}_{\pm}\right|$
increases as the string parameter value decreases. Analogously, Fig.
\ref{Fig-DeltaE(Omega3)}(b) shows that the difference between energy
levels decreases as $n$ increases (for a fixed value of the string
parameter). The pile-up effect of the levels in the
energy spectrum as $n$ increases is also observed in 
the hydrogen atom, where the energy levels become closer for higher values of the
 principal quantum number $n$.

We already mentioned in Section \ref{subsec:MagField} that
non-commutativity can be physically interpreted as a magnetic field.
In the present section, we discussed that the non-commutative term engenders
the energy difference $\Delta\mathcal{E}_{\pm}$. In analogy with the
 quantum mechanics of the hydrogen atom, it is known that
the energy spectrum is affected by the presence of an external magnetic
field through the Zeeman effect \cite{Griffiths}, which is responsible
for lifting some degeneracies. In this sense, with regards to the effect on the energy levels,
  there exists a parallel between non-commutativity for the KG oscillator
and the Zeeman effect for the hydrogen atom.
 Also regarding the effects on the
energy levels, one could affirm that there is an analogy between
the coupling $\omega L$ observed for the KG oscillator in Eqs.
(\ref{eq:E_WhittM_complex_variable}), (\ref{eq:E_commut}) and (\ref{eq:E_non-commut}), and
the spin-orbit coupling $\left\langle \mathbf{S}\cdot\mathbf{L}\right\rangle $
of the hydrogen atom.

We emphasize that the above analogies pertain exclusively
to changes in energy levels. The system that we deal with in this
work (i.e. a KG oscillator in a rotating frame with momentum non-commutativity
in a cosmic-string background space-time) is not of a physical nature similar
to the proton-electron system forming the hydrogen atom.




\section{Concluding remarks\label{Conclusion}}

This paper presents the study of a Klein-Gordon oscillator in a cosmic-string
background space endowed with a rotating frame and non-commutativity.

We introduced the oscillation through a non-mininal coupling with
the radial coordinate and the complex number, which introduces a change
in sign when the prescription operates onto the complex-conjugate
of the spin-zero field. This sign change between $\mathbf{p}\phi$
and $\mathbf{p}\phi^{*}$ is key for the physical interpretation of
the non-commutativity parameters. In Part 2 of this paper \cite{NCSF_Part2},
we shall explore a different coupling that leads to interpreting the
spin-zero case as the scalar sector of the Duffin-Kemmer-Petiau field
in the ways explored by Ref. \cite{Castro} in the commutative case.

We introduced non-commutativity via the generalized Bopp shift and
began with non-commutativity both in the coordinate space and in the
momentum space. Then we specified the Lagrangian of the system and
its equation of motion for non-commutativity restricted to momentum
space. This particular choice is based on firm physical grounds for,
as we have shown, the momentum-space non-commutativity is equivalent
to the presence of a magnetic field. Conversely, coordinate-space
non-commutativity does not map to a magnetic field. That is why we
decided to consider non-commutativity in momentum-space only.

Our study of the scalar field continued with the solution to the field
equations by applying the separation of variables to build the time-dependent
wave function. The space part of the latter was determined after we
restricted ourselves to non-commutativity in the momentum coordinate
pointing in the direction of the linear topological defect, i.e. by
assuming a uniform magnetic field directed along the string.

The time-independent wave function $\psi$ contains the ordinary periodic
angular solution with its associated quantum number $L=0,\pm1,\pm2,\dots$.
The $z$-dependent part of $\psi$ has a Gaussian form, shaped by
the mass $m$ of the oscillator and its natural frequency $\omega_{0}$;
the separation constant $K$ for this part of the solution is not
constrained or quantized by boundary conditions. The radial $\rho$-dependent
part of $\psi$ is not so simple to compute because the non-commutativity
parameter $\Omega_{3}$ appears, up to power two, in several terms
of the differential equation for $R\left(\rho\right)$. Our strategy
was to define the parameter $\epsilon=\frac{\Omega_{3}\omega}{2m\omega_{0}}$
and analyze the instances where it is small, i.e. where non-commutativity
($\Omega_{3}$) and frame rotation ($\omega$) are sub-dominant effects
when compared to $m\omega_{0}$. Accordingly, we have studied extensively
only the solution obtained from keeping linear terms in $\Omega_{3}$
in the differential equation for $R\left(\rho\right)$ and neglecting
terms scaling with $\epsilon$ and $\epsilon^{2}$. In this way, we
have obtained $R\left(\rho\right)$ in terms of Whittaker functions
multiplied by a Gaussian-type term $e^{\nicefrac{-\xi^{2}}{2}}$ and
the moderator factor $\xi^{\nicefrac{-1}{2}}$, where $\xi=\sqrt{m\omega_{0}}\rho$.

An important distance scale is the hard-wall position, at $\rho_{0}=1/\eta\omega$;
it sets the interval $\left[0,\rho_{0}\right]$ within which the particle
should be confined. This establishes a boundary condition leading
to energy quantization. In fact, the energy eigenvalues were shown
to scale as the inverse of the principal quantum number $n$ in the
case of a hard-wall position at infinity, $\rho_{0}\rightarrow\infty$,
a condition that is attained, for example, by slowing down the frame
rotation to the limit $\omega\rightarrow0$. Here the paramount role
of $\Omega_{3}$ is evident: the quantization is lost in the commutative
limit. It is also noteworthy that the non-relativistic limit is duly
obtained.

In the circumstances where the hard-wall takes on finite values, it
is necessary to perform a numerical analyzis of the zeros of the Wittaker
function to determine the particle energy levels. Here, the effect
of non-commutativity is to increase the absolute values of energy.
We emphasize the coupling between $\Omega_{3}$ and $\omega$; for
$L=0$, the eigenvalues reduce to their commutative counterparts.

All the above remarks regarding the time-independent wave function
$\psi$ and its energy eigenvalues $\mathcal{E}$ are valid when we
use the distance variable $\xi=\sqrt{m\omega_{0}}\rho$ to integrate
the differential equation for the radial part. The alternative definition
$\xi=\left[\left(m\omega_{0}\right)^{2}+\left(\Omega_{3}\omega/2\right)^{2}\right]^{\nicefrac{1}{4}}$
includes non-commutativity in the very definition of distances and
leads to different conclusions; with this latter definition, non-commutativity
introduces energy dissipation. However, the importance of this interesting
result is questionable since there is no meaningful non-relativistic
limit in this approach, as it is shown in Appendix A.




\section*{Acknowledgements}

RRC is grateful to the Instituto Tecnol\'ogico de Aeron\'autica (SP, Brazil)
for its hospitality and to CNPq (grant 309984/2020-3) for financial
support. MdeM is grateful to the Natural Sciences and Engineering
Research Council (NSERC) of Canada for partial financial support (grant
number RGPIN-2016-04309), to the Instituto Tecnol\'ogico de Aeron\'autica 
(SP, Brazil) and the Universidade Federal de Alfenas, Campus Po\c cos
de Caldas (MG, Brazil) for their hospitality. We are grateful to the
reviewers of an earlier version of this paper for helpful suggestions.
We also thank the referee's of the enhanced version of the paper for
their insightful comments that led to improvements.




\global\long\def\theequation{A-\arabic{equation}}
 \setcounter{equation}{0} 

\section*{Appendix A: Energy eigenvalues in a dissipative context\label{sec:Appendix:Dissipation}}

As mentioned after Eq. \eqref{eq:S(epsilon)}, the solution obtained
with this definition leads to interesting results, like dissipative
quantized energy, but it also causes problems, which we will discuss
in this Appendix. First, let us return to and analyze Eq. (\ref{eq:DiffEqR})
which covers the radial dependence of the wave function utilizing
a different definition of variable that will encompass the non-commutative
parameter $\Omega_{3}$. If we introduce the following parameters
with dimension of energy, 
\begin{equation}
S=S\left(m,\omega_{0},\omega,\Omega_{3}\right)\equiv\left[\left(m\omega_{0}\right)^{2}+\left(\frac{\Omega_{3}}{2}\omega\right)^{2}\right]^{\frac{1}{4}},\label{eq:S}
\end{equation}
\begin{equation}
T=T\left(\mathcal{E},m,\omega_{0},\omega,\Omega_{3}\right)\equiv\left[\left({\cal E}+L\omega\right)^{2}-m^{2}+3m\omega_{0}-Km\omega_{0}-\frac{1}{\eta^{2}}\left(\frac{1}{2}\Omega_{3}\right)^{2}\right]^{1/2},\label{eq:T}
\end{equation}
and the dimensionless variable 
\begin{equation}
\xi\equiv S\rho,\label{eq:xi}
\end{equation}
we rewrite Eq. (\ref{eq:DiffEqR}) in the form: 
\begin{align}
0 & =\partial_{\xi}^{2}R\left(\xi\right)+\frac{1}{\xi}\partial_{\xi}R\left(\xi\right)+2\frac{m\omega_{0}}{S^{2}}\xi\partial_{\xi}R\left(\xi\right)+\xi^{2}R\left(\xi\right)+2\left(\frac{1}{2}\Omega_{3}\right)\frac{\left({\cal E}+L\omega\right)\omega}{S^{3}}\xi R\left(\xi\right)\nonumber \\
 & +\left(\frac{T}{S}\right)^{2}R\left(\xi\right)-\Omega_{3}\frac{L}{\eta^{2}\xi}\frac{1}{S}R\left(\xi\right)-\frac{L^{2}}{\eta^{2}\xi^{2}}R\left(\xi\right).\label{eq:DiffEqR(xi)}
\end{align}
The solution to this differential equation is 
\begin{equation}
R\left(\xi\right)=e^{-\frac{\left(\xi-a\right)^{2}}{2b^{2}}}\xi^{\frac{L}{\eta}}H_{B}\left(2\frac{L}{\eta},i\frac{2a}{b^{3}},-\left[\left(\frac{T}{S}\right)^{2}+\frac{a^{2}}{b^{4}}-\left(2\frac{L}{\eta}+4\right)\frac{m\omega_{0}}{S^{2}}\right]\frac{1}{b^{2}},-4i\left(\frac{1}{2}\Omega_{3}\right)\frac{L}{\eta^{2}}\frac{1}{S}\frac{1}{b},ib\xi\right),\label{eq:R(xi)}
\end{equation}
where $H_{B}$ is the biconfluent Heun function with the definitions
\begin{equation}
a=\left(\frac{1}{2}\Omega_{3}\right)\frac{\left({\cal E}+L\omega\right)\omega}{S^{3}\left[\left(\frac{m\omega_{0}}{S^{2}}\right)\frac{1}{b^{2}}-1\right]},\quad\text{and}\quad\frac{1}{b^{2}}=\frac{m\omega_{0}}{S^{2}}\pm\sqrt{\left(\frac{m\omega_{0}}{S^{2}}\right)^{2}-1}.\label{eq:(a,b,c)}
\end{equation}

In order to caracterize our physical system, besides the wave function,
we need to study the energy and its eventual quantization. Our goal
is to treat this problem analytically although in an approximate approach.
Accordingly, let us consider the case where the non-commutativity
and the rotation of the frame are subdominant effects with respect
to the oscillator's mass and the oscillator's frequency, i.e. $\Omega_{3}\omega\ll m\omega_{0}$.
In this case, 
\begin{equation}
b\simeq\left(-1\right)^{j}\left\{ 1\mp\frac{i}{2^{3/2}}\left(\frac{1}{2}\frac{\Omega_{3}\omega}{m\omega_{0}}\right)+\frac{1}{2^{4}}\left(\frac{1}{2}\frac{\Omega_{3}\omega}{m\omega_{0}}\right)^{2}\right\} ,\qquad j=1,2.
\end{equation}
Note that the parameter $b$ is a complex number that exhibits real
and imaginary parts. This leads to a complex variable for the KG-oscillator
described by the Heun function. Moreover, under the approximation
$\Omega_{3}\omega\ll m\omega_{0}$: 
\begin{equation}
a\simeq\mp2^{1/2}i\frac{\left({\cal E}+L\omega\right)}{\left(m\omega_{0}\right)^{1/2}}\left[1\mp2^{1/2}i\left(\frac{1}{2}\frac{\Omega_{3}\omega}{m\omega_{0}}\right)-\frac{13}{8}\left(\frac{1}{2}\frac{\Omega_{3}\omega}{m\omega_{0}}\right)^{2}+\ldots\right].
\end{equation}

We are interested in the polynomial version of the Heun function in
the radial solution. This will be addressed in the following.

\subsection*{A.1. Energy for the hard-wall in the limit $\rho_{0}\rightarrow\infty$}

Heun functions $H_{B}\left(\alpha,\beta,\gamma,\delta,x\right)$ can
be cast in a polynomial form under the conditions \cite{Arriola}:
\begin{equation}
\gamma-2-\alpha=2n,\quad n=0,1,2,\ldots\qquad{\rm and}\qquad A_{n+1}=0\label{eq:PolynomialCond}
\end{equation}
From a physical perspective, this is necessary in order to avoid divergences.
The numbers $A_{n+1}$ are the series coefficients which obey the
recurrence relation: 
\begin{equation}
A_{l+2}-A_{l+1}\left\{ \left(l+1\right)\beta+\frac{1}{2}\left[\delta+\beta\left(1+\alpha\right)\right]\right\} +A_{l}\left(\gamma-2-\alpha-2l\right)\left(l+1\right)\left(l+1-\alpha\right)=0,\quad l\ge0,
\end{equation}
with 
\begin{equation}
A_{0}=1\qquad{\rm and}\qquad A_{1}=\frac{1}{2}\left[\delta+\beta\left(1+\alpha\right)\right].
\end{equation}
Our Eq. (\ref{eq:PolynomialCond}) leads to 
\begin{align}
{\cal E}_{\pm} & =\pm\left\{ m\omega_{0}\left(2n+2+\frac{2L}{\eta}\right)\left[1-\left(8\pm5\right)\frac{i}{\sqrt{2}}\left(\frac{1}{2}\frac{\Omega_{3}\omega}{m\omega_{0}}\right)+\left(29\pm28\right)\left(\frac{1}{2}\frac{\Omega_{3}\omega}{m\omega_{0}}\right)^{2}\right]-\left(\frac{m\omega_{0}}{\omega\eta}\right)^{2}\left(\frac{1}{2}\frac{\Omega_{3}\omega}{m\omega_{0}}\right)^{2}\right. \nonumber \\
 & \left.-\left[m^{2}+m\omega_{0}\left(1+K+\frac{2L}{\eta}\right)\right]\left[1-\sqrt{2}i\left(4\pm2\right)\left(\frac{1}{2}\frac{\Omega_{3}\omega}{m\omega_{0}}\right)+\frac{1}{2}\left(61\pm64\right)\left(\frac{1}{2}\frac{\Omega_{3}\omega}{m\omega_{0}}\right)^{2}\right]\right\} ^{1/2}-L\omega.
\end{align}
For consistency, we expand this expression in powers of $\frac{1}{2}\frac{\Omega_{3}\omega}{m\omega_{0}}$
up to second order, which results in 
\begin{align}
{\cal E}_{\pm} & \approx-\omega L\pm\sqrt{m\omega_{0}\left(2n+1-K\right)-m^{2}}\nonumber \\
 & \pm\frac{1}{2}\sqrt{2}i\frac{m\omega_{0}\left[K\left(4\pm2\right)-\left(4\pm3\right)-n\left(8\pm5\right)\mp\frac{L}{\eta}\right]+m^{2}\left(4\pm2\right)}{\sqrt{m\omega_{0}\left(2n+1-K\right)-m^{2}}}\frac{1}{2}\frac{\Omega_{3}\omega}{m\omega_{0}}\nonumber \\
 & \pm\frac{1}{2}\left[\frac{\left\{ m\omega_{0}\left[n\left(58\pm56\right)+\left(\frac{55}{2}\pm24\right)-K\left(\frac{61}{2}\pm32\right)-\frac{L}{\eta}\left(3\pm8\right)\right]-\left(\frac{m\omega_{0}}{\omega\eta}\right)^{2}-m^{2}\left(\frac{61}{2}\pm32\right)\right\} }{\sqrt{m\omega_{0}\left(2n+1-K\right)-m^{2}}}\right.\nonumber \\
 & \left.+\frac{1}{2}\frac{\left\{ m\omega_{0}\left[K\left(4\pm2\right)-\left(4\pm3\right)-n\left(8\pm5\right)\mp\frac{L}{\eta}\right]+m^{2}\left(4\pm2\right)\right\} ^{2}}{\left[m\omega_{0}\left(2n+1-K\right)-m^{2}\right]^{3/2}}\right]\left(\frac{1}{2}\frac{\Omega_{3}\omega}{m\omega_{0}}\right)^{2}\nonumber \\
 & \mp\frac{1}{2}\frac{\left(\frac{1}{2}\frac{\Omega_{3}}{\eta}\right)^{2}}{\sqrt{m\omega_{0}\left(2n+1-K\right)-m^{2}}}.\label{eq:E-O(2)}
\end{align}
The term in the second line contains a complex number, provided the
argument of the square root in its denominator is positive; that is,
if $m\omega_{0}\left(2n+1-K\right)-m^{2}>0$. The presence of the
complex number in Eq. (\ref{eq:E-O(2)}) can be interpreted as a dissipative
term; in fact, $\phi=e^{-i{\cal E}t}\psi=e^{-i\text{Re}\left(\mathcal{E}\right)t}e^{\pm\text{Im}\left(\mathcal{E}\right)t}\psi$.
The positive sign in the factor containing $\text{Im}\left(\mathcal{E}\right)$
leads to the divergence of the wave function; it is thus an unacceptable
choice. The negative sign in the factor containing $\text{Im}\left(\mathcal{E}\right)$
attenuates the wave function as the time increases. Therefore, this
implies that the half-life of the particle associated to the wave-function
is related to the expansion parameter $\frac{1}{2}\frac{\Omega_{3}\omega}{m\omega_{0}}$.

Note that the condition $m\omega_{0}\left(2n+1-K\right)-m^{2}>0$,
implies an upper bound on the dimensionless constant $K$ defined
in Eq. \eqref{eq:zeta_K}, namely, 
\begin{equation}
K<2n+1-\frac{m}{\omega_{0}}.\label{eq:constraint_K}
\end{equation}
This constraint is not arbitrary; in fact, it is mandatory if we expect
no dissipation to be observed in the commutative limit.

We notice that the complex number is associated to a coupling between
$\Omega_{3}$ and $\omega$. We could eliminate the complex dissipative
term in two ways: (1) with $\Omega_{3}=0$, which leads to discard
all the information on non-commutativity, or (2) with $\omega=0$,
which eliminates frame rotation but keeps the effect of non-commutativity
in the last term of Eq. (\ref{eq:E-O(2)}).

The global sign in the term of the second line of Eq. (\ref{eq:E-O(2)})
depends on the signs of the several physical parameters in it. In
order to see that more clearly, we write down Eq. (\ref{eq:E-O(2)})
in its two possible forms.

\subsubsection*{Energy eigenvalue ${\cal E}_{+}$}

With the upper sign in Eq. (\ref{eq:E-O(2)}), a physical solution
requires the second line of Eq. (\ref{eq:E-O(2)}) to satisfy 
\begin{equation}
\left[\omega_{0}\left(6K-7-13n-\frac{L}{\eta}\right)+6m\right]\left(\frac{1}{2}\frac{\Omega_{3}\omega}{m\omega_{0}}\right)\le0\qquad\left(\text{convergence condition}\right).\label{eq:converg_cond_E_plus}
\end{equation}
If the term in the second line of \eqref{eq:E-O(2)} is zero, we obtain
\begin{equation}
\frac{m\omega_{0}\left[6K-7-13n-\frac{L}{\eta}\right]+6m^{2}}{\sqrt{m\omega_{0}\left(2n+1-K\right)-m^{2}}}\left(\frac{1}{2}\frac{\Omega_{3}\omega}{m\omega_{0}}\right)=0,
\end{equation}
then the energy dissipation does not occur. The trivial ways to achieve
this condition, as pointed out above, are (1) in the commutative context,
$\Omega_{3}=0$, and (2) for a non-rotating frame, $\omega=0$. Alternatively,
this imposes the condition 
\begin{equation}
K=\frac{1}{6}\left(7+13n+\frac{L}{\eta}\right)-\frac{m}{\omega_{0}}\qquad\left(\text{conservative system}\right)
\end{equation}
upon the separation constant $K$. For this particular value of $K$,
the contribution of non-commutativity to the energy is of the second
order only -- see the three last lines in \eqref{eq:E-O(2)}. Moreover,
this value of $K$ constrains $L$ to assume negative values with
$\left|L/\eta\right|>n+1$ in order to guarantee that the condition
$m\omega_{0}\left(2n+1-K\right)-m^{2}>0$ is satisfied.

The condition for convergence of the solution, Eq. (\ref{eq:converg_cond_E_plus}),
allows for the following possibilities: (1) $\Omega_{3}\omega>0$
(non-commutativity and frame rotation in the same direction) and (2)
$\Omega_{3}\omega<0$ (non-commutativity and frame rotation in opposite
directions). The first possibility implies: 
\begin{equation}
K<\left(2n+1-\frac{m}{\omega_{0}}\right)+\frac{1}{6}\left(n+1+\frac{L}{\eta}\right),\qquad\left(\Omega_{3}\omega>0\right).
\end{equation}
This condition is already satisfied for $n+1+L/\eta>0$. Otherwise,
it is more restrictive than condition (\ref{eq:constraint_K}). The
second possibility requires 
\begin{equation}
K>\left(2n+1-\frac{m}{\omega_{0}}\right)+\frac{1}{6}\left(n+1+\frac{L}{\eta}\right),\qquad\left(\Omega_{3}\omega<0\right),
\end{equation}
which, together with (\ref{eq:constraint_K}), demands $\left(L/\eta\right)<-\left(n+1\right)$.

\subsubsection*{Energy eigenvalue ${\cal E}_{-}$}

If we select the lower sign in Eq. (\ref{eq:E-O(2)}), the second
line in ${\cal E}_{-}$ should comply with 
\begin{equation}
\left[m\omega_{0}\left(2K-1-3n+\frac{L}{\eta}\right)+2m^{2}\right]\left(\frac{1}{2}\frac{\Omega_{3}\omega}{m\omega_{0}}\right)\ge0,\qquad\left(\text{convergence condition}\right),\label{eq:converg_cond_E_minus}
\end{equation}
in order to avoid divergences of the wave function. If $\Omega_{3}\neq0$
and $\omega\neq0$, the equality is attained for 
\begin{equation}
K=-\frac{1}{2}\left(n+1+\frac{L}{\eta}\right)+2n+1-\frac{m}{\omega_{0}},\qquad\left(\text{conservative system}\right),
\end{equation}
which concurs with (\ref{eq:constraint_K}) under the requirement
$\frac{L}{\eta}>-\left(n+1\right)$. This would correspond to a system
without dissipation of energy. When we admit the possibility for dissipation,
the convergence condition opens two possibilities: first, for $\Omega_{3}\omega>0$,
the relation (\ref{eq:converg_cond_E_minus}) leads to 
\begin{equation}
K>-\frac{1}{2}\left(n+1+\frac{L}{\eta}\right)+2n+1-\frac{m}{\omega_{0}},\qquad\left(\Omega_{3}\omega>0\right)
\end{equation}
which, together with Eq. (\ref{eq:constraint_K}), leads to the same
conclusion as before: $\frac{L}{\eta}>-\left(n+1\right)$. Second,
for $\Omega_{3}\omega<0$, the inequality (\ref{eq:converg_cond_E_minus})
gives 
\begin{equation}
K<-\frac{1}{2}\left(n+1+\frac{L}{\eta}\right)+2n+1-\frac{m}{\omega_{0}},\qquad\left(\Omega_{3}\omega<0\right).
\end{equation}
This is automatically satisfied under Eq. (\ref{eq:constraint_K})
for $\frac{L}{\eta}<-\left(n+1\right)$. Otherwise, the inequality
above is more restrictive than (\ref{eq:constraint_K}).

When we study the limits of $\mathcal{E}$ related to particular regimes
of the physical parameters $\left\{ \omega_{0},\omega,\Omega_{3}\right\} $
there appears a caveat, namely, that the non-relativistic limit leads
to an imaginary rest mass. The root of this problem resides in the
definition of the radial variable in Eqs. (\ref{eq:xi}) and (\ref{eq:S}),
which includes the non-commutative parameter $\Omega_{3}$. This is
the reason why we preferred the radial variable $\xi=\sqrt{m\omega_{0}}\rho$
defined in Eqs. \eqref{eq:xi(S,rho)} and \eqref{eq:S(m,omega_0)}
of Section \ref{subsec:Solution}.

\section*{Appendix B: solution of the radial equation up to order $\epsilon^{1}$\label{sec:Appendix:epsilon1}}

In this appendix we discuss in details the difficulties encountered
when we keep terms up to first power in $\epsilon=\frac{\Omega_{3}\omega}{2m\omega_{0}}$
in the radial part of the field equation, Eq. (\ref{eq:DiffEqR(xi,epsilon)}).

We write Eq. \eqref{eq:DiffEqR_epsilons} back in terms of $\Omega_{3}$
and neglect the terms scaling with $\left(\Omega_{3}\right)^{2}$:
\begin{equation}
0=R^{\prime\prime}\left(\xi\right)+\frac{1}{\xi}R^{\prime}\left(\xi\right)+2\xi R^{\prime}\left(\xi\right)+\xi^{2}R\left(\xi\right)+A\xi R\left(\xi\right)+BR\left(\xi\right)-C\frac{1}{\xi}R\left(\xi\right)-D\frac{1}{\xi^{2}}R\left(\xi\right).\label{ApB1}
\end{equation}
where 
\begin{equation}
\begin{split} & A\equiv2\frac{\left({\cal E}+L\omega\right)}{\sqrt{m\omega_{0}}}\left(\frac{1}{2}\frac{\Omega_{3}\omega}{m\omega_{0}}\right),\\
 & B\equiv\frac{1}{\left(m\omega_{0}\right)}\left[\left({\cal E}+L\omega\right)^{2}-m^{2}+\left(3-K\right)m\omega_{0}\right],\\
 & C\equiv\frac{2}{\sqrt{m\omega_{0}}}\left(\frac{1}{2}\frac{\Omega_{3}}{\eta}\right)\left(\frac{L}{\eta}\right),\\
 & D\equiv\left(\frac{L}{\eta}\right)^{2}.
\end{split}
\end{equation}

In order to try and solve the previous equation via the Frobenius
method, we propose the ansatz 
\begin{equation}
R\left(\xi\right)=F\left(\xi\right)e^{-\frac{1}{2}\xi^{2}}e^{\sqrt{2-B}\xi}\xi^{\frac{L}{\eta}}.\label{RxiFxi}
\end{equation}
Now we replace this result in Eq. \eqref{ApB1} and search solutions
by means of the following series expansion for $F\left(\xi\right)$:
\begin{equation}
F\left(\xi\right)=\sum_{j=0}^{\infty}c_{j}\xi^{j+r}.
\end{equation}
Then by substitution into Eq. \eqref{ApB1} and rearrangement according
to powers of $\xi$, it follows that 
\begin{equation}
\begin{split} & c_{0}r\left(r+2\frac{L}{\eta}\right)=0,\\
 & c_{1}\left[\left(1+r\right)\left(1+r+2\frac{L}{\eta}\right)\right]+c_{0}\left[2\sqrt{2-B}r+\left(2\frac{L}{\eta}+1\right)\sqrt{2-B}-C\right]=0,\\
 & c_{2}\left[\left(2+r\right)\left(2+r+2\frac{L}{\eta}\right)\right]+c_{1}\left[2\sqrt{2-B}\left(1+r\right)+\left(2\frac{L}{\eta}+1\right)\sqrt{2-B}-C\right]=0,\\
 & c_{j+3}\left[\left(j+3+r\right)\left(j+3+r+2\frac{L}{\eta}\right)\right]+c_{j+2}\left[2\sqrt{2-B}\left(j+2+r\right)+\left(2\frac{L}{\eta}+1\right)\sqrt{2-B}-C\right]+Ac_{j}=0.
\end{split}
\end{equation}
The series is truncated if there is a maximum value $n$ for the index
$j$ such that the coefficient multiplying $c_{j+2}$ vanishes, which
leads to 
\begin{equation}
{\cal E}=-L\omega\pm\sqrt{m^{2}-\left(1-K\right)m\omega_{0}-\left(\frac{1}{2}\frac{\Omega_{3}}{\eta}\right)^{2}\left(\frac{L}{\eta}\right)^{2}\frac{1}{\left[n+2+r+\frac{L}{\eta}+\frac{1}{2}\right]^{2}}}.\label{eq:E_quantized_epsilon1}
\end{equation}
Moreover, additional constraints upon other coefficients must be imposed:
(1) In order to get $c_{n+3}=0$, it is necessary to demand that $c_{n}=0$;
(2) for $c_{n+4}=0$, we need $c_{n+1}=0$; and (3) a $c_{n+5}=0$
already guarantees the truncation of the series but at the cost of
requiring $c_{n+2}=0$. However, if $c_{n+2}=0$, then we do not need
to kill its coefficient in the series, which ultimately eliminates
the energy quantization requirement \eqref{eq:E_quantized_epsilon1}.

We could still argue that the energy quantization is valid together
with the imposition $c_{n+2}=0$. Then, we would still have to demand
that $c_{n}=c_{n+1}=0$; this amounts to at least four conditions
to impose. However, we have only three separation constants $\left\{ \mathcal{E},L,K\right\} $
to be constrained. On top of this, the consistency of the series truncation
should be assessed; this could lead to the undesirable demand for
a greater number of constraints, an additional possible caveat that
we will not investigate any further.

\bigskip{}


\end{document}